\documentclass[letterpaper]{article}

\usepackage[T1]{fontenc}

\usepackage{geometry}
\usepackage{setspace}

\usepackage[style = chem-acs]{biblatex}
\usepackage{graphicx}
\usepackage{float}
\newfloat{scheme}{htbp}{los}
\floatname{scheme}{Scheme}
\floatname{chart}{Chart}
\newfloat{graph}{htbp}{loh}

\usepackage{chemformula} 
\usepackage[version=4]{mhchem} 
\usepackage{amsmath}
\usepackage{amssymb}
\usepackage{amsmath}
\usepackage{authblk}
\author[1]{Aldo S. Pasos-Trejo} 
\author[2]{Daniel P. Ramirez-Echemendia} 
\author[3,4]{Daniel Conde-Torres}
\author[2]{D. Peter Tieleman}
\author[1]{Cecilia Clementi*}
\affil[1]{Department of Physics, Freie Universität Berlin, Arnimallee 12, 14195, Berlin, Germany}
\affil[2]{Centre for Molecular Simulation and Department of Biological Sciences, University of Calgary, 2500 University Drive NW, Calgary, Alberta T2N 1N4, Canada}
\affil[3]{Departamento de Física Aplicada, Facultade de Física, Universidade de Santiago de Compostela, Campus Vida, Santiago de Compostela, E-15782 A Coruña, Spain}
\affil[4]{Organic Chemistry Department, Centro Singular de Investigación en Química Biolóxica e Materiais Moleculares (CiQUS), Universidade de Santiago de Compostela, Campus Vida, Santiago de Compostela, E-15782 A Coru\~na, Spain}

\newcommand*\papertitle{Exploring the World of Lipids with Machine-Learned Coarse-Grained Models} 
\title{\papertitle}
\date{*Email:cecilia.clementi@fu-berlin.de}

\begin{document}

\maketitle

\begin{abstract}

Lipid bilayers are central to cellular organization and function, yet their
atomistic modelling remains constrained by its high computational cost. While
coarse-grained (CG) models can overcome this bottleneck, they are limited in
accuracy due to the use of simplified functional forms based on molecular
mechanics. Using a hybrid functional form combining physics-based terms with a
graph neural network, we present implicit-solvent, machine-learned CG models for
POPC. When trained on single-lipid data, the model
recovers the atomistic conformational landscape. By adding short atomistic
simulations of a small flat bilayer to the training data and combining
force matching with a denoising objective, we train a model that accurately
reproduces bilayer structure and dynamics, in closer agreement with the
atomistic reference than existing models, like Martini3 and DryMartini. Beyond its training data, 
the model stabilizes curved geometries, including vesicles, and captures complex 
cooperative phenomena such as bilayer and vesicle self-assembly and pore sealing.
While limitations remain, particularly regarding chemical transferability, our
results establish machine-learned coarse-grained potentials as a promising path
towards robust and accurate CG models of lipids and membranes.
\end{abstract}

\section{Introduction}
Biological membranes define the boundaries of cells and organelles while
actively regulating cellular processes such as molecular transport, signaling,
and energy transduction~\cite{sezginMysteryMembraneOrganization2017}. Molecular
dynamics (MD) simulations have proven to be a powerful tool for investigating
these systems, as they provide a atomic-level picture of membrane
organization and dynamics that is difficult to access
experimentally~\cite{marrinkComputationalModelingRealistic2019}. However, many
membrane phenomena occur over system sizes and timescales that are out of reach
for fully atomistic representations. This has driven decades of development of
coarse-grained (CG) lipid
models~\cite{marrinkMARTINIForceField2007,pedersenMartini3Lipidome2025}, which
trade molecular detail for computational efficiency, increasing the
spatiotemporal window accessible to
simulations~\cite{ingolfssonPowerCoarseGraining2014}.
\par
The Martini CG force field (FF) ranks among the most popular for membrane
simulations~\cite{marrinkTwoDecadesMartini2023}. Martini lipid models map, on
average, four heavy atoms onto a single bead,
resulting in an efficient trade-off between chemical specificity and
computational cost~\cite{marrinkPerspectiveMartiniModel2013}. It is important,
however, to distinguish the mapping itself from the strategy used to describe
the effective interactions among beads. In its standard formulation, Martini
relies on conventional molecular-mechanics terms where pairwise contributions
dominate~\cite{marrinkMARTINIForceField2007,souzaMartini3General2021}. While
effective for many applications, this functional form is not only difficult to
parametrize but also constitutes a strong approximation to the CG potential
of mean force (PMF) required in a bottom-up coarse-graining approach for the CG model to be thermodynamically consistent with a reference atomistic model~\cite{noidPerspectiveAdvancesChallenges2023}. The thermodynamically consistent PMF is many-body in nature and incorporates
solvent-mediated and entropic
contributions~\cite{jinUnderstandingMissingEntropy2019}. The resulting
limitations in accuracy and transferability follow as a direct consequence of
this simplification~\cite{noidPerspectiveCoarsegrainedModels2013}.
\par
Recently, machine-learned CG (MLCG) potentials have been proposed as a promising alternative
to molecular-mechanics
approximations for proteins~\cite{wangMachineLearningCoarseGrained2019,husicCoarseGrainingMolecular2020,charronNavigatingProteinLandscapes2025}. Instead of assuming
a fixed pairwise functional form, these bottom-up MLCG models learn the effective interaction energy among the beads as a function of the local environment, through training on atomistic
trajectories. Among these approaches, graph neural network (GNN)
architectures, in which CG beads serve as nodes of an interaction graph and
forces are learned from local geometric and chemical descriptors, have enabled
transferable, bottom-up coarse graining of protein systems~\cite{charronNavigatingProteinLandscapes2025}. Crucially,
these architectures preserve the translation, rotation, and permutation
symmetries of the underlying configurations, while naturally capturing many-body
correlations through successive message-passing
steps~\cite{husicCoarseGrainingMolecular2020,durumericMachineLearnedCoarsegrained2023}.
\par
A rigorous foundation for these methods is provided by the variational force-matching formulation~\cite{noidMultiscaleCoarsegrainingMethod2008}, which states that the functional form for the CG energy that minimizes
the mean squared error between predicted CG forces and forces mapped from an
atomistic trajectory recovers the exact many-body CG PMF in the infinite-data
limit~\cite{vothCoarseGrainingCondensedPhase2008}. At the same time, new training schemes that combine standard force
matching with noise-based objectives unify bottom-up coarse graining with ideas
from generative modeling, improving data efficiency and robustness when only
limited data are
available~\cite{durumericLearningDataefficientCoarsegrained2026}. A number of
variants of these models, most notably those derived from the CGSchNet
architecture~\cite{husicCoarseGrainingMolecular2020}, have demonstrated that CG
GNN potentials can capture folding free-energy landscapes of small proteins with
higher accuracy than traditional CG
FFs~\cite{majewskiMachineLearningCoarsegrained2023,charronNavigatingProteinLandscapes2025}.
These applications have so far remained largely
confined to proteins. Here we extend the MLCG approach to lipid membranes.
\par 
Membrane systems pose their own challenges. Lipid bilayers are dense,
two-dimensional liquids in which the relevant physics is governed by collective
packing and by the largely entropic hydrophobic effect, rather than by a set of
metastable chain conformations. This point is particularly important because
water molecules, which CG GNN architectures typically treat implicitly,
provide the dominant driving force holding the bilayer
together~\cite{israelachviliModelPackingLipids1975}. Early efforts have shown
that GNN-based CG lipid models can be built from atomistic
references~\cite{liaoDevelopmentCoarseGrainedLipid2025}, but they have so far
relied on an ultra-CG resolution ($\simeq$ six beads per lipid), which obscures
the link to the underlying chemistry. To our knowledge, no assessment of GNN
potentials for lipid bilayers at the Martini resolution, benchmarked against
both the atomistic data and the Martini model, has yet been reported.
\par
To fill this gap, we used the MLCG
framework~\cite{charronNavigatingProteinLandscapes2025} to train a CG GNN
potential for 1-palmitoyl-2-oleoyl-sn-glycero-3-phosphocholine (POPC) at the
resolution of the Martini~3 lipidome mapping~\cite{pedersenMartini3Lipidome2025},
with solvent effects included implicitly. The model was trained only on
atomistic simulations of a single solvated lipid and of a flat bilayer, and was
then tested on systems of increasing complexity. For a single lipid, it
reproduces the atomistic intramolecular geometry and conformational free-energy
landscape. For flat bilayers, it
reproduces the atomistic bilayer structure, thickness and acyl-chain order more
accurately than both Martini~3 and DryMartini~\cite{arnarez2015dry}. Without any additional training
data, the model also keeps vesicles and micelles stable, drives the spontaneous
self-assembly of bilayers and vesicles from disordered configurations, and seals
membrane pores. Finally, we test its chemical transferability to DOPC and DPPC
bilayers, an assessment that reveals the current limitations of the approach.

\section{Methods}\label{sec:methods} 
\begin{figure}[t!]
    \centering
    \includegraphics[width=\textwidth]{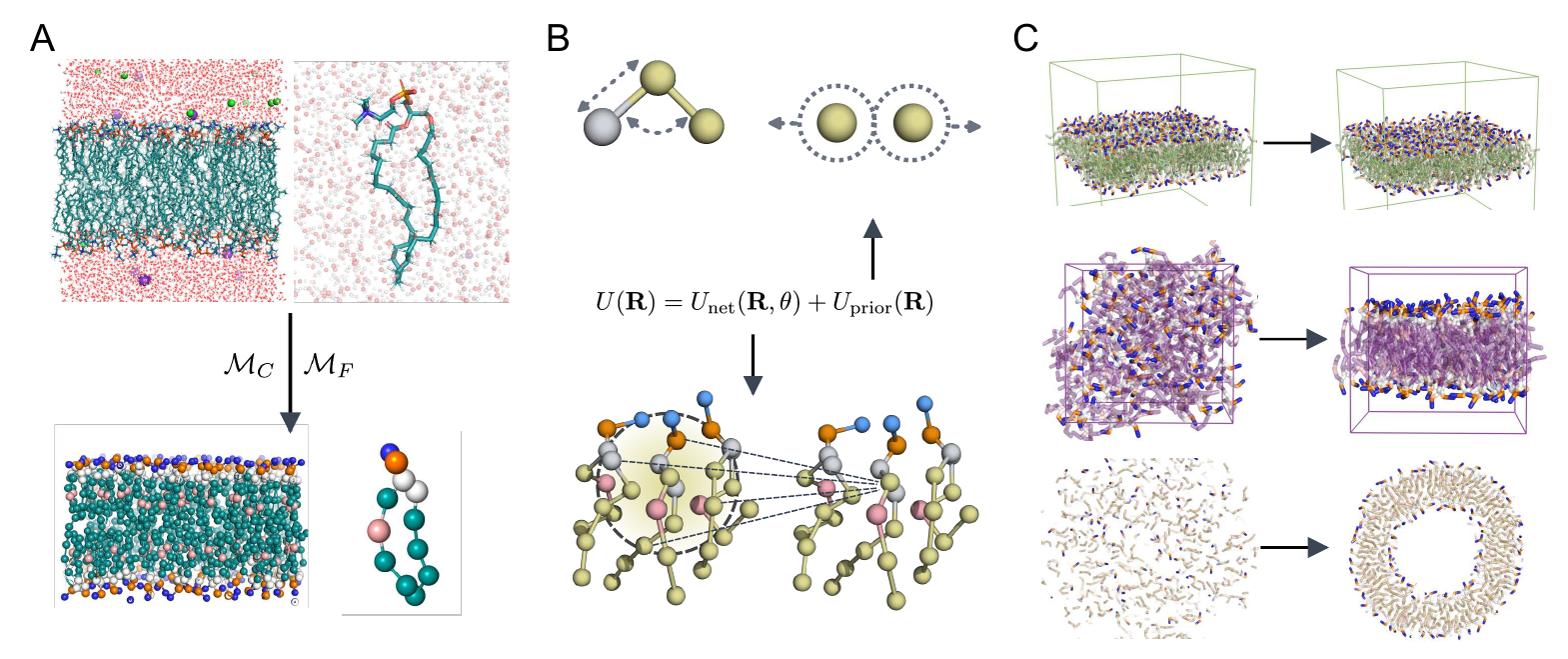}
    \caption[Overview of the machine-learned coarse-graining pipeline.]{Overview
    of the machine-learned coarse-graining pipeline.            
        (A) A Martini~3 CG mapping $\mathcal{M}_C$ and its corresponding force map $\mathcal{M}_F$ are used to 
        project the reference atomistic simulations of POPC.
        (B) A hybrid energy function $U$ with two parts is used: a prior, which enforces physical constraints, and an ML-based energy 
        that is learned from the reference atomistic simulations.
        (C) The energy function is learned and validated across different scenarios: bilayers, vesicles, and self-assembly.}
    \label{fig:mlcg_pipeline}
\end{figure}

The pipeline for constructing  lipid MLCG
models is illustrated in Figure~\ref{fig:mlcg_pipeline}. First, atomistic MD trajectories
of POPC were generated and mapped onto the twelve-bead-per-lipid Martini~3 CG
representation. A set of physics-inspired terms (the \textit{prior} energy) was then parametrized by Boltzmann
inversion of the distributions of the atomistic configurations in the training dataset mapped on the CG resolution. A graph neural network with the SchNet architecture
was trained to learn the remaining contribution to the CG potential. The combined prior $+$ SchNet
potential was then used to run CG Langevin dynamics simulations. 
\subsection{Dataset construction}
Using a conventional molecular dynamics setup (as detailed in section S1 of the Supplementary Information), we generated atomistic trajectories in the NVT ensemble at a temperature of 300~K for two systems: a single lipid in a small water box, which provides information on tail--water hydrophobic
interactions, and a fully assembled bilayer of 150 lipids, which provides conformations with the expected
bilayer ordering as well as head--water hydrophilic interactions. Table~\ref{tab:training-data} summarizes
the resulting training data. Details on the generation of the training data and the simulation parameters
are given in the Supporting Information.
\begin{table}[t]
    \centering
    \caption{Atomistic training data.}
    \label{tab:training-data}
    \begin{tabular}{cccc}
        \hline
        System            & Simulated time & Saving stride & Number of frames   \\
        \hline
        Single lipid      & 56.7 ns        & 40 fs         & 1.42$\times 10^6$  \\
        Assembled bilayer & 700 ns         & 20 ps         & 3.5$\times 10^4$  \\
        \hline
    \end{tabular}
\end{table}
\par
The atomistic trajectories were then mapped onto the Martini~3 lipidome
representation~\cite{pedersenMartini3Lipidome2025}, which consists of twelve
beads per lipid molecule: choline, phosphate, two glycerol beads, and eight
acyl-chain segments. Each CG bead is assigned a disjoint set of atoms
(Table~\ref{tab:popc-mapping}), and, for a given atomistic
configuration, the bead is placed at the center of geometry of its atoms. This
mapping is linear and can therefore be written explicitly as a matrix
$\mathcal{M}_C \in \mathbb{R}^{3N\times3n}$, where $n$ is the number of atoms in
the atomistic system and $N$ is the number of CG beads, so that the CG
coordinates $\mathbf{R}$ associated  to atomistic
coordinates $\mathbf{r}$ are obtained as $\mathbf{R} = \mathcal{M}_C(\mathbf{r})$. For the forces, we constructed an associated optimal linear force map
$\mathcal{M}_F$, which minimizes the variance of the mapped forces and correctly
accounts for the bond constraints in our atomistic
simulations~\cite{kramer_statistically_2023} (see Supporting Information for details).
\subsection{Model architecture}
For given coordinate and force maps $\mathcal{M}_C$ and $\mathcal{M}_F$, the MLCG framework~\cite{husicCoarseGrainingMolecular2020}
learns a hybrid physics/ML energy function of the form
\begin{equation}
    U(\mathbf{R},\theta) = U_{\text{net}}(\mathbf{R},\theta) + U_{\text{prior}}(\mathbf{R}) 
\end{equation}
where $U_{\text{net}}$ is a graph neural network and $U_{\text{prior}}$ is a physics-based term.
The prior energy, $U_{\text{prior}}$, comprises bond, angle and non-bonded terms:
\begin{align}
    U_{\text{prior}} &= U_{\text{bonds}} + U_{\text{angles}} + U_{\text{non-bonded}} \\
    &= \sum_{(i,j) \in\text{Bonds}} k_{ij}(l_{ij} - l_{ij}^{(0)} )^2 \\
    &+ \sum_{(i,j,k) \in\text{Angles}} \; \sum_{d=0}^4 \; k_{ijk}^{(d)} \phi_{ijk}^d \\
    &+ \sum_{(i,j) \in \text{non-bonded}} \text{Cut}\left(\frac{6}{\alpha_{ij}}\exp{\left[\alpha_{ij}\left(1-\frac{r_{ij}}{r_{ij}^{(0)}}\right) \right] }\right) 
\end{align}
with
\begin{equation}
    \text{Cut}(V(r_{ij})) = V(r_{ij}) - V(r_c) - V'(r_c) (r_{ij}-r_c)
\end{equation}
a linear correction that makes $U_{\text{non-bonded}}$ and the associated forces vanish
continuously at the cutoff radius $r_c$, which is the same 12~\AA{} cutoff used
by the network (see below). Here, $l_{ij}$ is the bond length, $\phi_{ijk}$ the
bond angle and $r_{ij}$ the distance between non-bonded beads. The non-bonded
sum runs over all pairs of beads within $r_c$, excluding bonded pairs or an angle triplet. The quantities $k_{ij}$,
$l_{ij}^{(0)}$, $k_{ijk}^{(d)}$, $\alpha_{ij}$ and $r_{ij}^{(0)}$ are parameters
fitted for each combination of bead types. The non-bonded term is purely repulsive
and represents excluded-volume interactions. The relevance of each prior term was
assessed through an ablation study (see Section~\ref{sec:si-sensitivity} in the Supporting Information).
\par
For the ML contribution $U_{\text{net}}(\mathbf{R},\theta)$, we use a message-passing GNN.
For a given configuration, a latent representation is initialized for every CG bead from its bead
type. This representation is then updated through several interaction blocks, each of which
aggregates learned ``messages'' computed from the latent representations of the neighboring beads
and the distances between the bead and each neighbor. Specifically, we use the SchNet
architecture~\cite{schuttSchNetDeepLearning2018}, which is fast to evaluate and has been used
extensively in the MLCG literature~\cite{husicCoarseGrainingMolecular2020,charronNavigatingProteinLandscapes2025,durumericLearningDataefficientCoarsegrained2026}.
\par
Because of the large number of beads per configuration and the high density of the system, the
network cutoff strongly affects the size of the message-passing graph and, consequently, the
required GPU memory, particularly during training. We therefore chose a cutoff of 12~\AA{} based on
the radial distribution function of the beads, as this distance covers the first two peaks,
corresponding to the first and second neighboring shells (Figure~\ref{fig:RDF}). Testing
models with other cutoff values (see Section~\ref{sec:si-sensitivity} in the Supporting Information) supports this choice. The
remaining network hyperparameters are listed in Table~\ref{tab:hyperparams} (Section~~\ref{si:hyperparams}).
\subsection{Prior fitting, model training and validation}
We parametrized $U_{\text{prior}}$ via Boltzmann inversion: for every prior term
and combination of CG bead types, we converted the distribution of the associated
feature into a free energy and fitted the parameters of the corresponding
functional form (by least squares for the bond and angle terms, and by a genetic
algorithm for the non-bonded term, see Section~\ref{si:prior-fitting}). The ML contribution
$U_{\text{net}}(\mathbf{R},\theta)$ was then learned with a gradient-based
optimizer using a generalized force-matching loss~\cite{durumericLearningDataefficientCoarsegrained2026}
\begin{equation}
    \mathcal{L}(\theta,\mathbf{T})= \left\langle \left\Vert -\nabla U (\mathcal{M}_C(\mathbf{r}),\theta) - \mathbf{T}\left(\mathbf{f}(\mathbf{r})\right) \right\Vert^2 \right\rangle_{\mathbf{r}}
\end{equation}
with 
\begin{equation}
    \mathbf{f}(\mathbf{r}) = \underbrace{\mathcal{M}_F (\mathbf{F}(\mathbf{r}))}_{\text{Atomistic forces}} + \underbrace{\nabla \log{\kappa}(\mathcal{M}_C(\mathbf{r}),\mathbf{r})}_{\text{Denoising contributions}}
\end{equation}
where $\kappa$ is a Gaussian noise kernel that adds noise to the coordinates and
$\mathbf{T}$ is an additional linear map that optimally combines the atomistic
force and denoising contributions~\cite{kramer_statistically_2023}. 
Since perturbing the 
mapped coordinates exposes the network to configurations slightly off those sampled in the 
atomistic trajectories, the denoising term regularizes the learned potential and improves both data
efficiency and stability in
simulation~\cite{durumericLearningDataefficientCoarsegrained2026}.
We use
Gaussian noise with zero mean and a standard deviation of $\sigma=0.001$~\AA{}
(see Section~\ref{sec:si-sensitivity} in the Supporting Information). Note that for $\sigma=0$ the
generalized loss reduces to the conventional force-matching loss used in previous
MLCG work~\cite{husicCoarseGrainingMolecular2020,charronNavigatingProteinLandscapes2025}.
\par
We trained two models, which differ in their training data, noise level and
number of interaction blocks (Table~\ref{tab:hyperparams}).
\texttt{SinLip-MLCG} is trained only on the single-lipid data, with the conventional force-matching loss, and
serves as a first test of whether the framework recovers the conformational
landscape of an isolated lipid. \texttt{Lipid-MLCG}, the model used for all
membrane simulations in this work, is trained on both the single-lipid and the
bilayer data, with a noise value ($\sigma=0.001$~\AA{}). Training proceeds over
minibatches, and because the two training systems differ
greatly in size, we fixed the composition of each \texttt{Lipid-MLCG}
minibatch to 8 bilayer and 160 single-lipid frames per GPU, so that both
systems contribute to every update (see Section~\ref{sec:si-sensitivity}).
\par
The force-matching validation loss is only indirectly related to a
model's ability to reproduce the PMF in
simulation~\cite{thaler2022deep,kramer_statistically_2023}, and this mismatch
becomes more pronounced with noise-augmented training, where the validation loss
is also considerably noisier. For the model trained without noise
(\texttt{SinLip-MLCG}), we selected as optimal epoch 140, which lies in the plateau 
region of the validation loss. For the models trained with noise, we instead trained 
each models for 100 epochs and then simulated several checkpoints to select the best one. 
After a comparison, we selected epoch 44 as it belongs to the region where the validation
loss began to plateau and appeared to best reproduce the reference simulations of the 
lipid bilayer (Fig. \ref{fig:epoch_selection}).

The selected model was validated by running CG molecular dynamics
simulations in the NVT ensemble with a Langevin thermostat (see Supporting
Information for details on the CG simulations). Bilayer simulations used periodic
boundary conditions with the minimum image convention.
\section{Results}
\subsection{Single lipid conformational landscape}
As an initial test, we evaluated whether the \texttt{SinLip-MLCG} model, trained
on the single-lipid trajectories alone without noise (see Methods), reproduces the atomistic behavior.  
We refer to this model as \texttt{SinLip-MLCG}. After training for $\sim 160$ epochs,
we selected epoch 140 in the plateau region of the validation loss as the optimal
model (Figure~\ref{fig:single_lipid_train_curve}). We then simulated
the \texttt{SinLip-MLCG} model and compared the results with the atomistic
reference and with a prior-only simulation.
\par
\begin{figure}[t!]
    \centering
    \includegraphics[width=\textwidth]{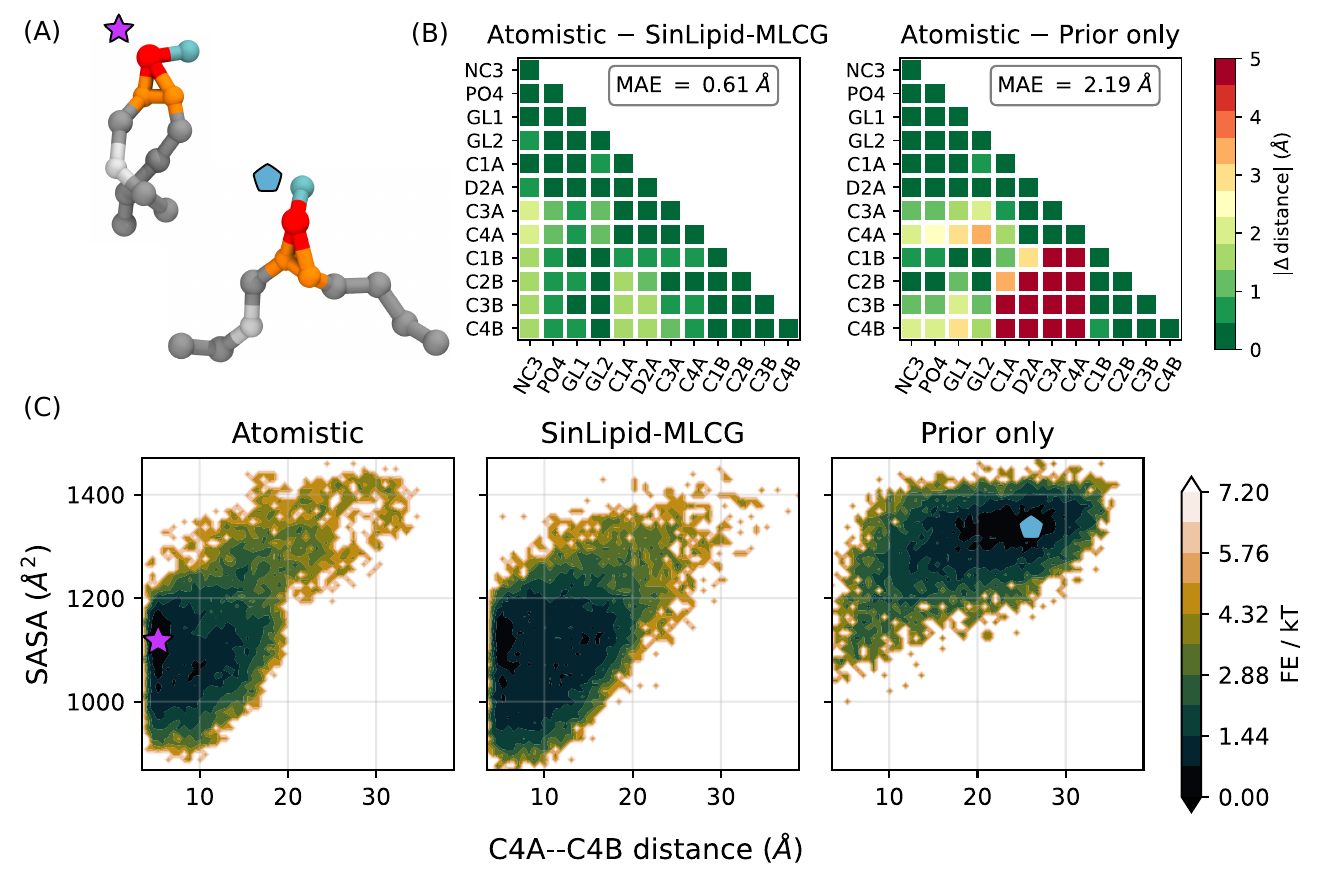}
    \caption[Evaluation of the SinLip-MLCG model.]{Evaluation of the
    \texttt{SinLip-MLCG} model, trained on single-POPC trajectory data.
    (A)~POPC structures corresponding to the metastable states in panel C.
    (B)~Lower-triangular pair-distance deviation matrices over the twelve CG
    beads of POPC, for the MLCG model (left) and the prior-only (right)
    simulations, relative to the mapped atomistic reference. (C)~Ensemble
    projections on the inter-tail-end distance $r_{\text{C4A--C4B}}$ and the
    solvent-accessible surface area (SASA), for the atomistic reference, MLCG
    and prior-only simulation (left to right). Cartoon insets show the
    projection of the configurations from the identified basins.}
    \label{fig:single_lipid}
\end{figure}

The results are shown in Figure~\ref{fig:single_lipid}. In panel B, we benchmark
the intramolecular geometry of each model with a pair-distance metric, in which
the mean intramolecular pair distances are compared with those of the mapped
atomistic reference~\cite{ramirez-echemendiaGradientBasedOptimizationForce2025}.
The prior-only simulation deviates strongly from the reference for the
inter-tail bead pairs of the acyl chains, whereas the MLCG model accurately
matches the mean distance for every pair.
\par
The quality of the MLCG model is further confirmed by the free energy surface
(FES) projected onto two reaction coordinates: the distance between the tail-end
beads C4A and C4B, and the solvent-accessible surface area (SASA)
(Figure~\ref{fig:single_lipid}C). These coordinates distinguish compact,
solvent-avoiding conformations from extended, solvent-exposed ones (examples in
Figure~\ref{fig:single_lipid}A). The FES of the MLCG model closely matches the
atomistic reference, with both displaying a basin corresponding to a compact
conformation of the lipid.
\subsection{Bilayer structure and dynamics} \label{sec:bilayer}
\begin{figure}[h!]
    \centering
    \includegraphics[width=\textwidth]{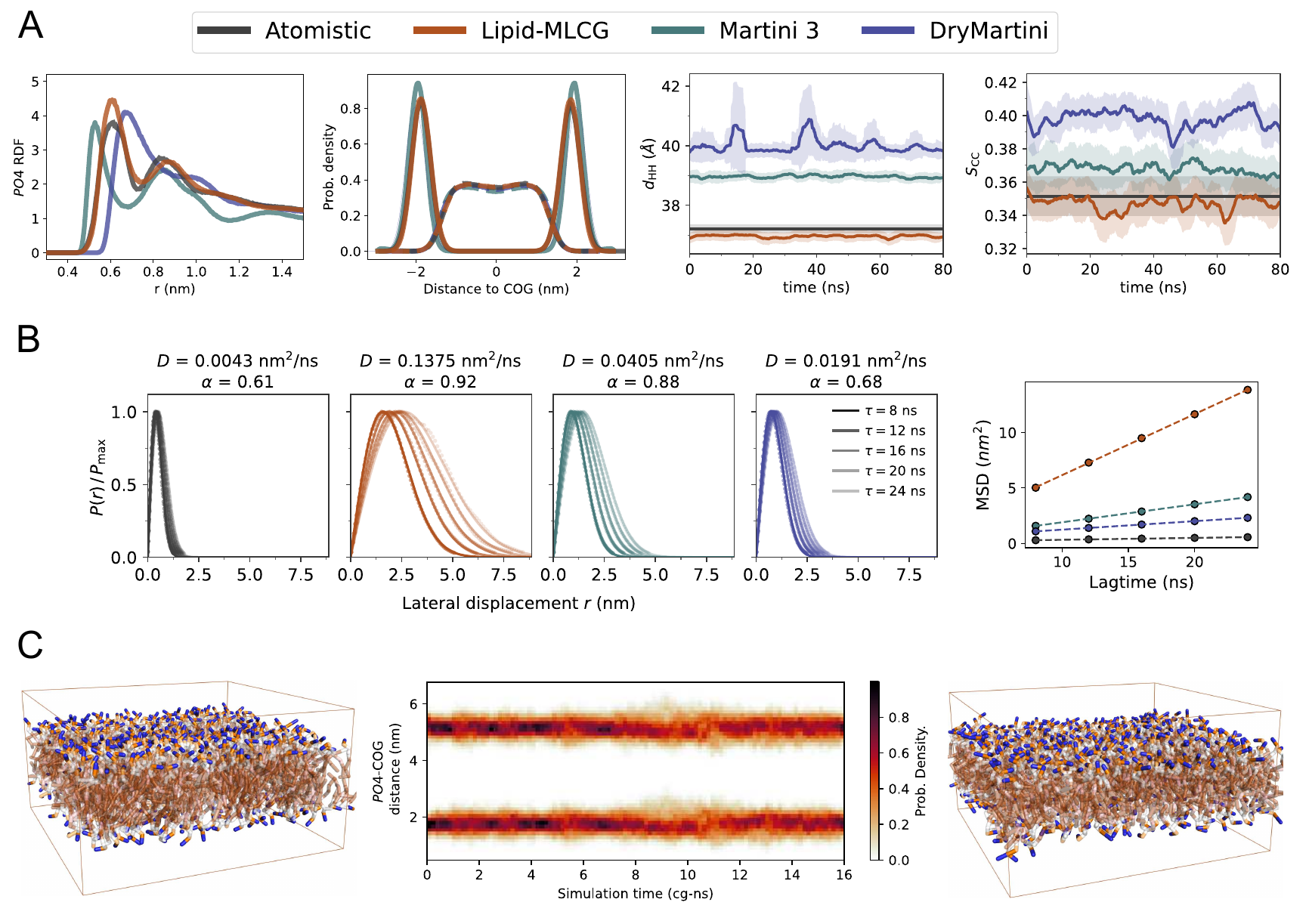}
    \caption[Bilayer structure and dynamics]{Performance of the
    \texttt{Lipid-MLCG} model for a 150-lipid bilayer. (A) Equilibrium and
    time-dependent properties. Left: RDF of PO4 headgroups, and $z$-density of
    PO4 headgroups (solid) and tail beads (dashed) for the equilibrium ensemble.
    Right: headgroup--headgroup distance $d_{HH}$ and tail order parameter
    $S_{CC}$ as a function of time. (B) Lateral displacement distributions and
    estimated diffusion constant $D$. (C) Heatmap of the PO4 $z$-density as a
    function of simulation time for a 600-lipid bilayer, with the initial and
    final configurations shown on either side.}
    \label{fig:bilayer}
\end{figure}

Having established that the \texttt{SinLip-MLCG} model recovers the behavior of
a single lipid, we trained and evaluated the \texttt{Lipid-MLCG} model on the full
dataset, which combines the single-lipid and the assembled-bilayer trajectories.
The main results are shown in Figure~\ref{fig:bilayer}, together with a
comparison with the explicit-solvent Martini~3~\cite{souzaMartini3General2021, pedersenMartini3Lipidome2025} and the
implicit-solvent DryMartini~\cite{arnarez2015dry} models (simulation details in the
Supporting Information). The \texttt{Lipid-MLCG} model reproduces equilibrium
properties of the atomistic ensemble, such as the radial distribution function
(RDF), more accurately than the other two models. In addition, its $z$-density profiles
closely match the atomistic reference, indicating that the model has learned the
correct shape of the membrane. Time-dependent structural properties, namely the
membrane thickness, measured as the PO4 headgroup--headgroup distance $d_{HH}$,
and the tail order parameter $S_{CC}$, remain stable during the simulation and match to the atomistic reference, while  the other two models present deviations
(Table~\ref{tab:bilayer-metrics}). The \texttt{Lipid-MLCG} model reproduces the
atomistic membrane thickness and the order parameter within the statistical error,
whereas both Martini models produce thicker and more ordered bilayers, with
$d_{HH}$ larger by $\sim$2~\AA{} (Martini~3) and $\sim$3~\AA{} (DryMartini). We also computed the 
orientational order parameter $P_2$ of the lipids, which uses a single head-to-tail vector 
per lipid and therefore measures the alignment of the molecule as a whole with respect to the bilayer normal, rather than the conformational order of the acyl chains measured by $S_{CC}$ 
(see Supporting Information). All four models yield a similar value, $P_2 \approx 0.79$--$0.80$, indicating that  the lipids are equally well aligned with the normal in every model.
\begin{table}[htbp]
  \centering
  \small
  \caption{Structural and dynamical properties of the 150-lipid POPC bilayer.
           Bilayer thickness $d_{HH}$, tail order parameter $S_{CC}$ and lipid
           orientational order $P_2$ are averaged over 80~ns (mean $\pm$
           standard deviation over frames). The lateral diffusion coefficient
           $D$ and the anomalous exponent $\alpha$
           ($\langle r^2 \rangle \sim \tau^{\alpha}$) are obtained from the
           lateral displacements of the PO4 beads at lag times of 8--24~ns.}
  \label{tab:bilayer-metrics}
  \begin{tabular}{lcccccc}
    \hline
    Model & $d_{HH}$ (\AA{}) & $S_{CC}$ & $P_2$ & $D$ ($10^{-3}$~nm$^2$/ns) & $D/D_{\text{atomistic}}$ & $\alpha$ \\
    \hline
    Atomistic           & $37.2 \pm 0.2$ & $0.35 \pm 0.01$ & $0.79 \pm 0.02$ & 4.3   & 1.0  & 0.61 \\
    \texttt{Lipid-MLCG} & $37.0 \pm 0.2$ & $0.35 \pm 0.01$ & $0.79 \pm 0.02$ & 137.5 & 32.1 & 0.92 \\
    Martini~3           & $39.0 \pm 0.2$ & $0.37 \pm 0.01$ & $0.79 \pm 0.01$ & 40.5  & 9.5  & 0.88 \\
    DryMartini          & $40.0 \pm 0.6$ & $0.40 \pm 0.01$ & $0.80 \pm 0.01$ & 19.1  & 4.5  & 0.68 \\
    \hline
  \end{tabular}
\end{table}

\par
Beyond these structural properties, we measured the lateral displacement of the
lipids in the bilayer. From the distribution of lateral displacements at five different 
lag times between 8 and 24~ns (Figure~\ref{fig:bilayer}B), we estimate a
diffusion coefficient for our model that is $\sim 32$ times larger than that of
the atomistic reference and $\sim 3-7$ times larger than those of the Martini
models (Table~\ref{tab:bilayer-metrics}). This indicates that the
dynamics of our CG model are considerably faster, as expected. The value of the anomalous diffusion exponent differs
between models: the \texttt{Lipid-MLCG} and Martini~3  simulations are close to 
the Fickian regime over this window, whereas the atomistic and DryMartini 
simulations are more subdiffusive, making the comparison between them 
more complex.
\par
To further test the stability of the model and its transferability to larger
systems, we simulated a 600-lipid bilayer, four times larger than the bilayer in
the training data. The model keeps the large membrane stable throughout the
simulation (Figure~\ref{fig:bilayer}C). We also probed the behavior of the model
in other layer-like systems. A monolayer initially assembles into the correct
shape, but eventually becomes unstable and splits into a bilayer with a
micelle-like edge (Figure~\ref{fig:monolayer}). Similarly, when extra 
padding space is added along one of the simulation box dimensions to break the layer continuity across the periodic boundary, the bilayer edge along that
dimension assumes a curved and micelle-like shape, while maintaining the linear shape along the other dimension (Figure~\ref{fig:wrongboundary}).
Having confirmed that our model stabilizes the bilayer, we assessed its design
choices by ablating the prior terms and varying the key training hyperparameters
(see Section~\ref{sec:si-sensitivity} in the Supporting Information). The results 
support our choices of prior terms and cutoff and, more
importantly, show that models trained with different random seed initializations reproduce the same behavior, demonstrating the robustness of the framework.
\subsection{Curved conformations}
\begin{figure}[h!]
    \centering
    \includegraphics[width=\textwidth]{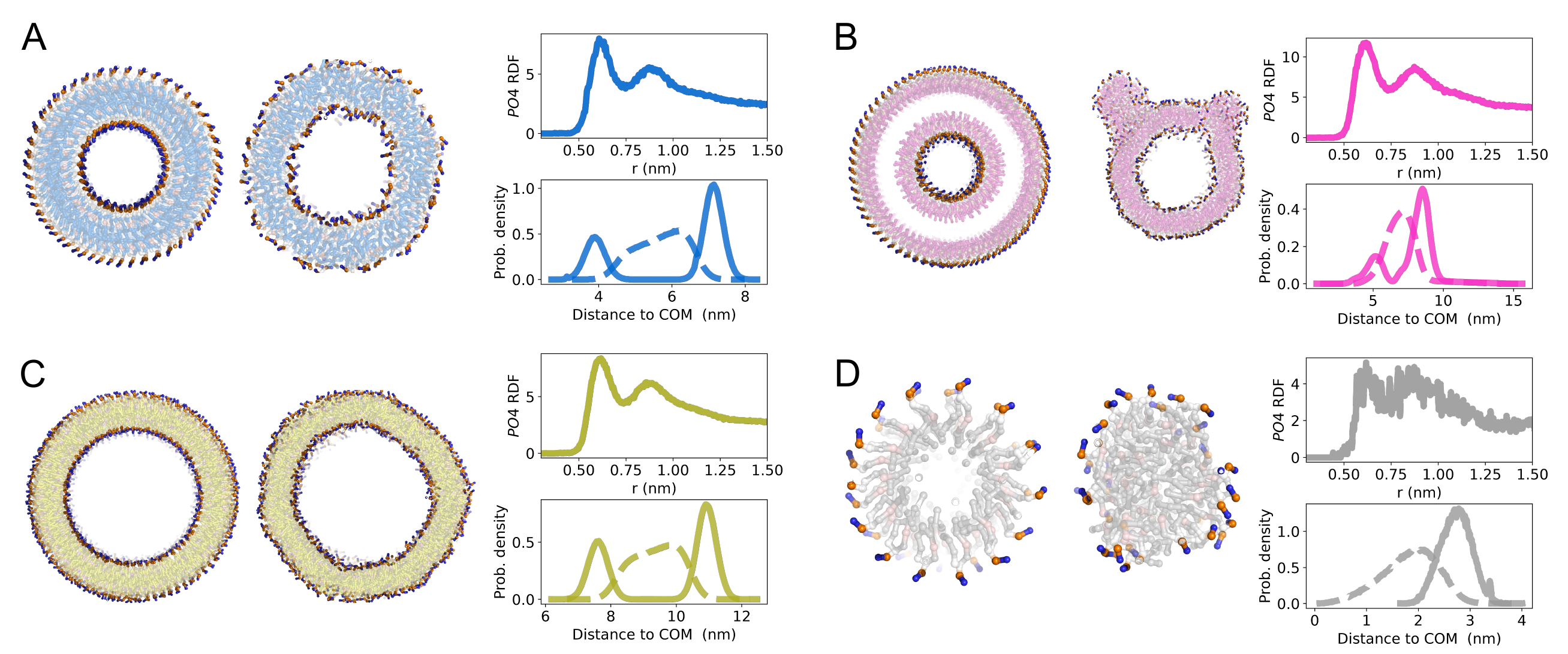}
    \caption[Curved conformations]{Performance of the \texttt{Lipid-MLCG} model
    on the curved systems described in Table~\ref{tab:curved-systems}: (A) small
    vesicle, (B)  vesicle initialized with an incorrect shape, (C) large vesicle and (D) spherical
    micelle. For each system, we show snapshots of the initial and final
    configurations of a 16~cg-ns trajectory, together with the RDF and the
    radial densities of head (solid) and tail (dashed) beads.}
    \label{fig:curved-confs}
\end{figure}

We next evaluated the model  performance on
curved, vesicle-like conformations. Curved environments provide a more stringent
test, as the training data contain no curved conformations; they are also
biologically relevant, as \textit{in vitro} studies typically make extensive use
of liposomes~\cite{matsuoka1998copii,peter2004bar}. We tested the lipid-MLCG model on four
curved systems, described in Table~\ref{tab:curved-systems}, including two preformed vesicles of
different size (small and large), a spherical micelle, and an incorrectly shaped vesicle.
The latter was constructed as a stress test, far from equilibrium, which has the same inner
radius as the small vesicle but a much larger outer radius. While the preformed systems test whether the model maintains a curved
structure, this system tests whether it can relax an unphysical one.
The results for these 
four systems are summarized in Figure~\ref{fig:curved-confs}. For the vesicles, 
the RDF of PO4 headgroups is similar to that of the flat bilayer, indicating appropriate 
local ordering, while for the micelle the distribution is noisier because of the smaller system
size. The radial densities of head and tail beads show the expected bilayer
thickness and a clear separation between the two groups, demonstrating that the
vesicles remain globular and stable throughout the simulation. This is further
confirmed by the time evolution of these densities and of the bilayer thickness
(Figures~\ref{fig:small_vesicle_time},
\ref{fig:large_vesicle_time}, \ref{fig:wrong_vesicle_time} and
\ref{fig:micell_time}).
\par
The dynamics of the vesicle initialized with an incorrect shape are of particular interest
(Figure~\ref{fig:wrong_vesicle_detail}). At the beginning of the simulation,
lipids from the inner leaflet expand to form a vesicle-like conformation. This
conformation is not stable, however, because of the strong imbalance between the
numbers of lipids in the inner and outer leaflets. To relieve the excess lateral
stress, the outer leaflet expels lipids, which aggregate into micelle-like
bulges that remain attached to it. This behavior indicates that the model
has learned the correct lipid density of a vesicle: excess lipids are
expelled until the remaining bilayer reaches that density.
\par
To quantify the stability of the curved systems, we computed the structural
metrics in Table~\ref{tab:curved-metrics}. After an initial relaxation, both
vesicles keep a stable size and remain nearly spherical, with a ratio between
their largest and smallest axes $ a_{\text{max}}/a_{\text{min}} \leq 1.08$.
However, curvature has an effect on the order of the lipid chains and the
bilayer thickness, reflected by a smaller $d_{HH}$, $S_{CC}$ and $P_2$ compared
to the flat bilayer. This curvature-induced disorder is also supported as the
deviation from the values of the flat bilayer increases with the curvature, from
the large vesicle, to the small vesicle and the micelle. When differentiating
between the inner and outer leaflet, and between headgroups and tails
(Table~\ref{tab:curved-leaflets}), we see that most of this disorder is in the
inner leaflet, whose tails point into a widening volume, while its headgroups
remain closely packed. Similar curvature-induced effects have been previously
reported in the literature~\cite{risseladaCurvatureEffectsLipid2009}. 
\begin{table}[htbp]
  \centering
  \caption{Curved systems used to test the model, in their initial
        configuration. Radii are the median distance from the center of mass of
        the aggregate to the phosphate headgroups of each leaflet (micelle: mean
        distance). $N_\text{out}/N_\text{in}$ is the number of lipids in the
        outer over the inner leaflet in the initial configuration.}
  \label{tab:curved-systems}
  \begin{tabular}{lcccc}
    \hline
    System &  $ R_\text{out} $ (nm) & $ R_\text{in}  $ (nm) & Number of lipids &
    $N_\text{out}/N_\text{in}$ \\
    \hline
    Small vesicle          & 6.68 & 3.14 & 959  & 2.10 \\
    Large vesicle          & 10.75 & 7.26  & 2758 & 1.58 \\
    Incorrectly shaped vesicle & 8.88 & 3.14 & 1546  & 4.00 \\
    Spherical micelle      & 3.39 & - & 60 & -- \\
    \hline
  \end{tabular}
\end{table}
\begin{table}[htbp]
    \centering
    \caption{Structure of the curved conformations simulated with
            \texttt{Lipid-MLCG}, compared with the flat bilayer
            (Table~\ref{tab:bilayer-metrics}). $R_\text{out}$ and $R_\text{in}$
            are the median PO4 distances of the outer and inner leaflets from
            the center of mass (or mean PO4 distance for the micelle), $d_{HH} =
            R_\text{out} - R_\text{in}$, and $S_{CC}$ is computed with respect
            to the local radial direction. $P_2$ is the orientational order of
            the whole-lipid director (PO4 to the midpoint of the tail ends) with
            respect to the radial direction at the PO4 bead. The area per lipid
            (APL) of each leaflet is $4\pi R^2/N$ at its PO4 radius. Values are
            the mean $\pm$ standard deviation over frames, excluding the first
            1~cg-ns, pooled over 4 (vesicles) or 10 (micelle) independent runs.}
        \label{tab:curved-metrics}
    \small
    \setlength{\tabcolsep}{4pt}
    \begin{tabular}{lccccccc}
        \hline
        System &  $\langle R_\text{out} \rangle$ (nm) & $\langle R_\text{in}
        \rangle $ (nm) & $a_\text{max}/a_\text{min}$ & $d_{HH}$ (\AA{}) &
        $S_{CC}$ & $P_2$ & APL in / out (nm$^2$) \\
        \hline
        Flat bilayer & -- & -- & -- & $37.0 \pm 0.2$ & $0.347 \pm 0.013$ & $0.79
        \pm 0.02$ & 0.673 / 0.673 \\
        Large vesicle & $10.95 \pm 0.02$ & $7.63 \pm 0.02$ & $1.04 \pm 0.01$ &
        $33.2 \pm 0.1$ & $0.268 \pm 0.003$ & $0.73 \pm 0.00$ & 0.683 / 0.892 \\
        Small vesicle & $7.15 \pm 0.02$ & $3.91 \pm 0.03$ & $1.08 \pm 0.03$ &
        $32.4 \pm 0.2$ & $0.255 \pm 0.006$ & $0.71 \pm 0.01$ & 0.623 / 0.989 \\
        Micelle & $2.72 \pm 0.02$ & -- & $1.17 \pm 0.05$ & -- & $0.116 \pm
        0.022$ & $0.63 \pm 0.04$ & -- / 1.555 \\
        \hline
    \end{tabular}

\end{table}

\subsection{Self-assembly and pore sealing}
We
next investigated whether the \texttt{Lipid-MLCG} model also captures more complex collective behavior,
such as bilayer self-assembly and pore sealing. Self-assembly is relevant for
biological processes~\cite{pohorille2009self} and has been observed in CG
simulations~\cite{marrink2001,marrinkMolecularDynamicsSimulation2003}, but
remains challenging for atomistic simulations~\cite{skjevik2015all}. Similarly,
pore formation and sealing are important membrane
processes~\cite{bischofberger2009membrane} that have been probed both
experimentally~\cite{ros2021techniques} and
computationally~\cite{kirsch2016membrane,zhang201983}.
\par
\begin{figure}[h!]
    \centering
    \includegraphics[width=\textwidth]{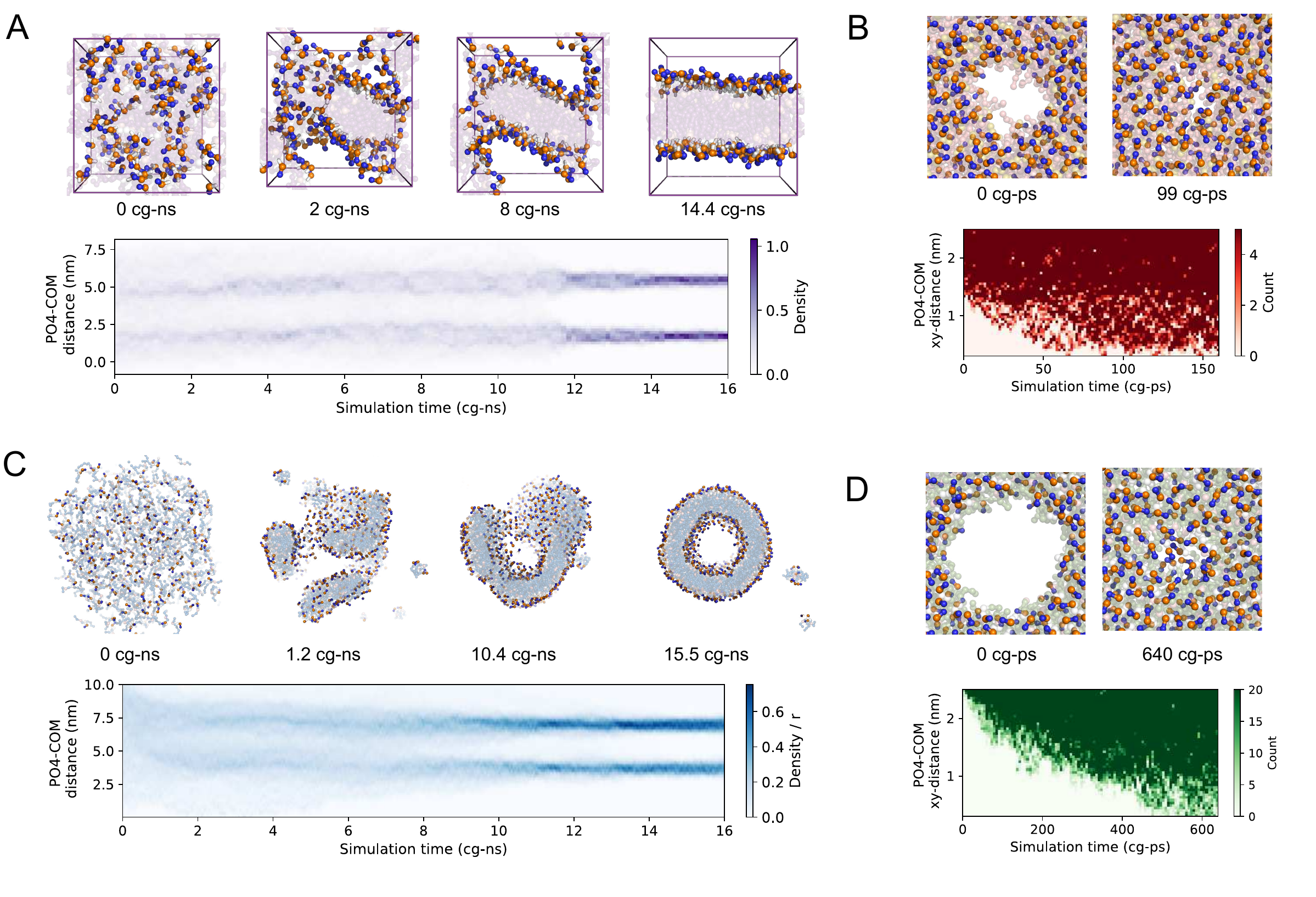}
    \caption[Self-assembly and pore sealing]{
        Self-assembly and pore-sealing simulations. (A, C) Bilayer and vesicle self-assembly: (top) snapshots of the simulation;
        (bottom) headgroup $z$-density or radial density as a function of time, showing the formation of a bilayer or vesicle.
        (B, D) Sealing of a small and a large circular pore: (top) snapshots of the simulation; (bottom) radial density as a
        function of simulation time, showing the pore closing.
    }
    \label{fig:complex-systems}
\end{figure}

We tested the model on four systems, described in
Table~\ref{tab:complex-systems}, with the results shown in
Figure~\ref{fig:complex-systems}. In the high-density system, the lipids
self-assemble into a bilayer: they first aggregate into micelles, which then
merge into a single bilayer-like structure that spans the periodic box. This
behavior is robust and was observed in four independent simulations
(Figure~\ref{fig:bilayer_self_assembly}). When the bilayer self-assembles on a plane normal to the $z$ direction
in the periodic box (Replica 3), it has a thickness
and chain order very close to the intact bilayer; when  it self-assemble along the $x$  
direction (Replica 4), the bilayer is $\sim$2~\AA{} thinner and less ordered in
each chain. Bilayers that self-assemble along a diagonal direction (Replicas 1 and 2)
 cover a larger area with the same number of lipids, and are therefore
thinner and less ordered, consistent with the trend observed in the  planar systems.
The self-assembly behavior is also observed in simulations with an initial low-density
configuration in which the system self-aggregates into a bicelle
(Figure~\ref{fig:bilayer_wrong_self_assembly}).
\par
A similar process occurs in a system with an initial configuration of medium-density, spherically distributed lipids:
the lipids first assemble into small independent micelles, which then come
together to form a stable vesicle. The self-assembled vesicles are very similar
to the preformed small vesicle with the same number of lipids
(Table~\ref{tab:complex-metrics} and Table~\ref{tab:vesicle-self-assembly}).
They are equally spherical and have comparable thickness and order, as shown by
$a_\text{max}/a_\text{min} $, $d_{HH}$ and $S_{CC}$, respectively. A small
difference in the radii $R_{\text{in}}$, $R_{\text{out}}$ can be explained
by the fact that not all lipids are incorporated into the vesicle, as only $97
\pm 2\%$ of the lipids form vesicle at the end of the simulation while
the remainder form a few small aggregates or remain in monomeric form.  This is also
related to a difference in the lipid distribution across the leaflets, indicated
by the discrepancy in the ratio between the number of lipids in each leaflet
$N_\text{out}/N_\text{in}$. This suggests that the leaflet composition of the
preformed vesicle is not the one preferred by the model, and that the
self-assembled vesicles relax to a more favorable distribution, with fewer
lipids in the strongly curved inner leaflet.  
The vesicle self-assembly was consistently replicated across two replicas for
each of four initial conditions (Figure~\ref{fig:vesicle_self_assembly}). 
\par
In the pore-sealing simulations, both the small and the large circular pores are sealed 
quickly, and the resulting bilayer remains stable and aligned, as shown by the $P_2$  values 
close to those of the intact bilayer (Table~\ref{tab:complex-metrics}). 
As the lipids spread out to fill the hole, the sealed bilayer ends 
up at a lower density (i.e. larger APL) than the intact one, and is therefore thinner 
and less ordered, shown by the smaller $d_{HH}$ and $S_{CC}$. This behavior is consistently observed 
in two independent trajectories for each system (Figure~\ref{fig:pore_closing}). 
\par
Together, these four tests provide strong evidence that the \texttt{Lipid-MLCG}
model has learned the correct amphipathic behavior of the lipids, as heads and
tails assemble in the right arrangement even from incorrectly shaped or disordered starting
configurations, such as the medium-density system. As with the curved systems,
these simulations show that the model extrapolates across conformational space
with the correct physical behavior, since its training data contain no
self-assembly or pore-sealing events.
\begin{table}[htbp]
  \centering
  \caption{Self-assembly and pore-sealing systems. The lipid density is the number of
    lipids over the volume of the periodic box. The pore radius is $\sqrt{A/\pi}$, with $A$
    the bead-free area of the pore in the bilayer plane in the initial configuration.}
  \label{tab:complex-systems}
  \begin{tabular}{p{3cm}p{5cm}cc}
    \hline
    System name & Description & Number of lipids & Lipid density (nm$^{-3}$) \\
    \hline
    High density, uniform distribution   & Lipids distributed uniformly at random inside a $7.0\times7.2\times7.6$~nm$^{3}$ rectangular box  & 150 & 0.391 \\
    Medium density, spherical distribution & Lipids positioned randomly in a roughly spherical distribution inside a $24\times24\times24$~nm$^{3}$ cubic box  & 959 & 0.069 \\
    Small pore & A $14\times14$~nm$^{2}$ bilayer with a circular pore of 1.06~nm radius & 570 & -- \\
    Large pore & A $14\times14$~nm$^{2}$ bilayer with a circular pore of 2.54~nm radius  & 514 & -- \\
    \hline
  \end{tabular}
\end{table}
\begin{table}[htbp]
  \centering
  \small
  \caption{Structure of the systems obtained by self-assembly and pore sealing with
    \texttt{Lipid-MLCG}, compared with the preformed reference systems (italics).
    For planar bilayers, the area per lipid (APL) is set by the box and the bilayer
    normal (Miller indices $(hkl)$, see Supporting Information); for vesicles, the
    APL is given for the inner / outer leaflet as $4\pi R^2/N$ at the PO4 radius of each
    leaflet (Table~\ref{tab:curved-metrics}), and
    $S_{CC}$ and $P_2$ are computed with respect to the local radial direction.
    $t$ is the assembly or sealing time. Values are the mean $\pm$ standard
    deviation over frames, pooled over runs: for the self-assembled and sealed systems,
    over the last 10\% of each run (16~cg-ns; 23--26~cg-ns for the vesicles); for the
    preformed references, which are at equilibrium, over the same frames as
    Tables~\ref{tab:bilayer-metrics} and~\ref{tab:curved-metrics}.
    $^\dagger$Computed against a local normal (2~nm neighbourhood)}
  \label{tab:complex-metrics}
  \begin{tabular}{lccccc}
    \hline
    System & APL (nm$^2$) & $d_{HH}$ (\AA{}) & $S_{CC}$ & $P_2$ & $t$ (cg-ns) \\
    \hline
    \multicolumn{6}{l}{\textbf{Bilayer self-assembly}} \\
    \textit{Intact bilayer}             & 0.673 & $37.0 \pm 0.2$ & $0.347 \pm 0.013$ & $0.79 \pm 0.02$ & -- \\
    Replica 3, normal $(001)$           & 0.673 & $37.0 \pm 0.2$ & $0.345 \pm 0.011$ & $0.79 \pm 0.01$ & 11.7 \\
    Replica 4, normal $(100)$           & 0.725 & $35.2 \pm 0.2$ & $0.307 \pm 0.013$ & $0.76 \pm 0.02$ & 9.0 \\
    Replica 1, normal $(01\bar{1})$     & 0.981 & $30.0 \pm 0.4$ & $0.198 \pm 0.017$ & $0.59 \pm 0.03$ & 10.2 \\
    Replica 2, normal $(01\bar{1})$     & 0.981 & $30.3 \pm 0.5$ & $0.202 \pm 0.015$ & $0.58 \pm 0.03$ & 13.2 \\
    \hline
    \multicolumn{6}{l}{\textbf{Vesicle self-assembly}} \\
    \textit{Preformed small vesicle}    & 0.623 / 0.989 & $32.4 \pm 0.2$ & $0.255 \pm 0.006$ & $0.71 \pm 0.01$ & -- \\
    Self-assembled vesicle (8 runs)     & 0.659 / 0.943 & $32.7 \pm 0.2$ & $0.269 \pm 0.008$ & $0.71 \pm 0.01$ & $6.9 \pm 3.4$ \\
    \hline
    \multicolumn{6}{l}{\textbf{Pore sealing}} \\
    Small pore, sealed                  & 0.709 & $35.2 \pm 0.2$ & $0.295 \pm 0.006^\dagger$ & $0.76 \pm 0.01^\dagger$ & $0.09 \pm 0.01$ \\
    Large pore, sealed                  & 0.786 & $33.4 \pm 0.1$ & $0.277 \pm 0.006^\dagger$ & $0.75 \pm 0.01^\dagger$ & $0.64 \pm 0.02$ \\
    \hline
  \end{tabular}
\end{table}

\subsection{Chemical transferability}
\begin{figure}[h!]
    \centering
    \includegraphics[width=\textwidth]{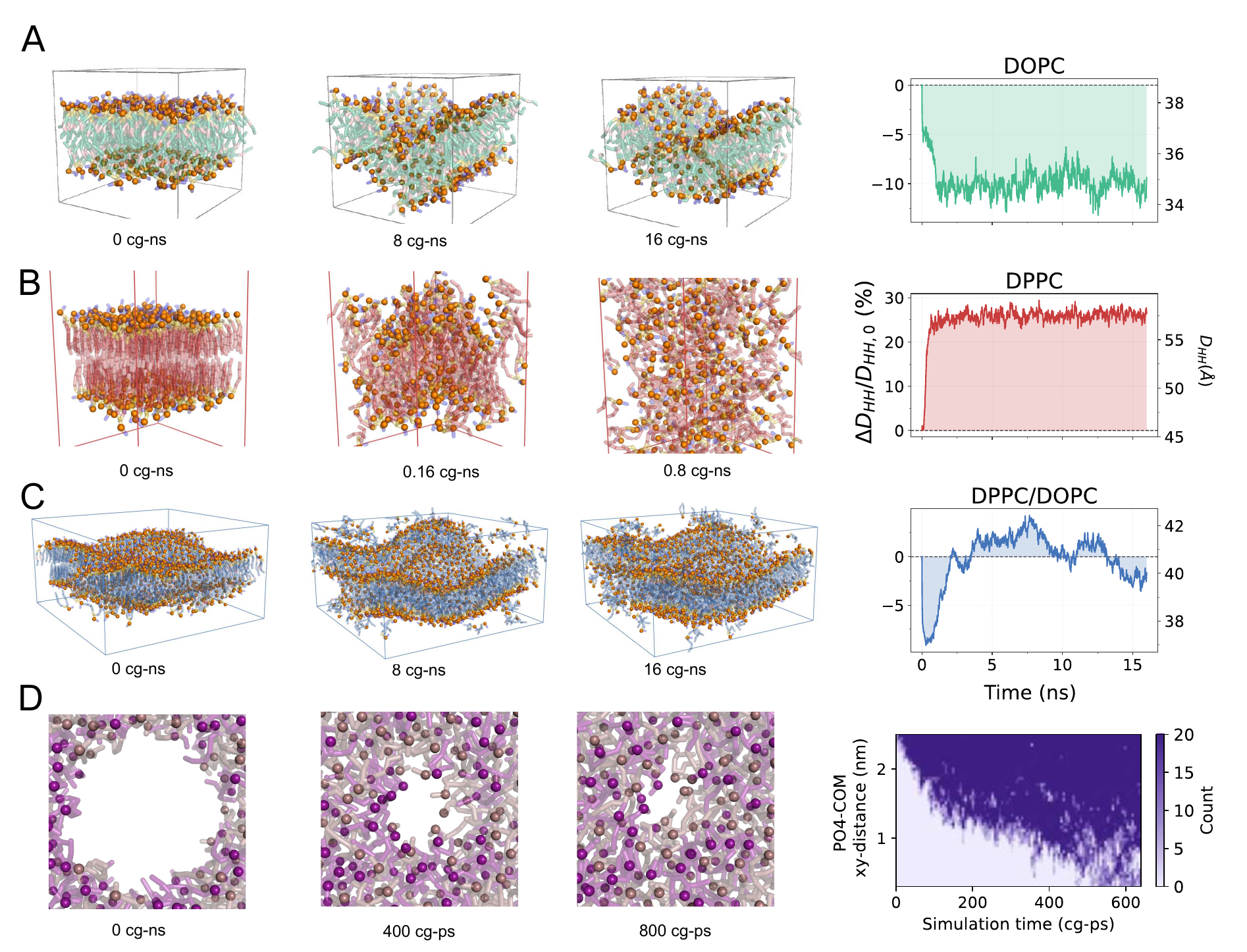}
    \caption[DOPC/DPPC simulations]{Simulations of DOPC/DPPC systems: (A) DOPC bilayer, (B) DPPC gel, (C) DOPC/DPPC bilayer and
    (D) DOPC/DPPC pore. Left: simulation snapshots. Right: bilayer thickness for (A)--(C) and radial density of headgroups for (D).}
    \label{fig:dppcdopc}
\end{figure}

Because POPC is asymmetric, with one saturated and one unsaturated tail, the
\texttt{Lipid-MLCG} model must learn the interactions of both chain types with
all other bead types. We therefore benchmarked its chemical transferability on
DOPC and DPPC, which carry two copies of the unsaturated (oleoyl) and the
saturated (palmitoyl) chain of POPC, respectively. We simulated four DOPC/DPPC
systems, described in Table~\ref{tab:dopcdppc}. The results are summarized in
Figure~\ref{fig:dppcdopc}.
\par
Although the model stabilizes a pure DOPC bilayer, the bilayer shows some
unexpected curvature. The mixed DOPC/DPPC bilayer behaves similarly and, in
addition, shows no sign of the DOPC/DPPC demixing into domains (Figure~\ref{fig:dppcdopc_no_aggr}) 
that is known to occur~\cite{schmidt_phase_2009}. The DPPC 
simulation fails to keep the system compact and quickly destabilizes
into a heavily disordered state. In contrast, the pore in the DOPC/DPPC mixture
seals as expected, and the resulting bilayer remains stable throughout the
simulation (Figure~\ref{fig:dppcdopc_hole_details}).
\par
Overall, the \texttt{Lipid-MLCG} model maintains the correct head--tail
arrangement in DOPC and DOPC/DPPC bilayers, but does not keep them perfectly
flat. The performance for pure DPPC deviates more strongly from the correct behavior. 
We note, however, that unlike POPC and DOPC, DPPC forms a gel phase at the simulated
temperature of 300~K. This phase is considerably denser than the
liquid-disordered phase sampled in our training data, and a model
parametrized at a different phase may not capture it. While the model 
extrapolates correctly to curved geometries and self-assembly processes, 
extrapolation across thermodynamic states and phases is considerably more 
challenging and is beyond the scope of the current design.
\begin{table}[htbp]
  \centering
  \caption{DOPC/DPPC systems. The pore radius is defined as in Table~\ref{tab:complex-systems}.}
  \label{tab:dopcdppc}
  \begin{tabular}{lp{4cm}ccc}
    \hline
    System & Description  & $N_{\text{DOPC}}$ & $N_{\text{DPPC}}$ & $N$ \\
    \hline
    DOPC bilayer      & DOPC bilayer  & 242 & -- & 242  \\
    DPPC gel          & DPPC gel & -- & 242 & 242 \\
    DOPC/DPPC bilayer & DOPC/DPPC bilayer & 460 & 692 & 1152  \\
    DOPC/DPPC pore    & DOPC/DPPC bilayer with a 2.55~nm radius pore  & 269 & 245 & 514 \\
    \hline
  \end{tabular}
\end{table}

\section{Discussion and Conclusions}
We have shown that MLCG potentials can be constructed for lipids at Martini resolution.
The \texttt{SinLip-MLCG} model describes a single lipid with high fidelity to the atomistic
reference, while the \texttt{Lipid-MLCG} model reproduces the structural properties of a lipid
bilayer in close agreement with the atomistic data. The latter also stably
simulates vesicles and micelles, preserving the separation between hydrophobic tails and
hydrophilic headgroups as well as the global ordering of the bilayer. Furthermore, it captures
complex cooperative processes such as bilayer and vesicle self-assembly and pore sealing, none
of which are present in its training set. Given that the model was trained on a small dataset
of under 1~\textmu s of aggregated simulation time, we consider these results remarkable:
the correct reproduction of collective processes absent from the training data suggests that the
model has learned the relevant physical interactions between CG beads, rather than merely fitting
the training configurations. Moreover, models trained with different random seeds consistently
stabilized the bilayer, which further supports the robustness of the framework.
\par
Nevertheless, the tests of chemical transferability reveal clear limitations. DOPC and DOPC/DPPC
systems form bilayer-like structures, but present a spurious curvature, and the DOPC/DPPC mixture
shows no sign of lipid demixing. The DPPC simulation is qualitatively incorrect, as the gel phase
is not captured. Part of these failures can be attributed to the mismatch between the
phase sampled in the training data (liquid-disordered) and those of the DPPC gel
and the DOPC/DPPC mixture. Because the learned CG potential is a potential of mean
force, and therefore thermodyamic state dependent, and applying it to a
different phase is itself an extrapolation,
in addition to the change in lipid chemistry.
We believe, however, that the limited transferability is also related to a known shortcoming of
the SchNet architecture: when trained on data from a single molecule, geometric message passing
causes distance information to dominate the latent representation of the beads at the expense of
chemical identity~\cite{pasostrejo2026exploration}. Chemical transferability with MLCG models is
achievable~\cite{charronNavigatingProteinLandscapes2025}, but it requires a chemically diverse
dataset and careful curation of the minibatch composition.
\par
Overall, this work demonstrates the feasibility of bottom-up MLCG lipid models at the
widely used Martini resolution, using a hybrid functional form that combines a graph neural
network with physics-based prior terms. Although MLCG models are typically more computationally
expensive than traditional force fields with simple functional forms, they are considerably more
expressive. We believe that this expressivity, in particular the message passing across local
neighborhoods of beads, is key to capturing the many-body effects arising from enthalpic, entropic
and solvent-mediated interactions, which pairwise CG potentials cannot accurately
represent~\cite{noidMultiscaleCoarsegrainingMethod2008,jinUnderstandingMissingEntropy2019}.
In addition, the dynamics of the \texttt{Lipid-MLCG} model are considerably faster than those of
both the atomistic reference and the state-of-the-art Martini~3 force field. As expected, 
the CG time is significantly accelerated, and its interpretation requires rescaling. However, it allows
efficient sampling of rare events and slow lipid processes beyond the timescales accessible to
atomistic MD.
\par
The combination of enhanced expressivity and accelerated dynamics makes the MLCG framework a
promising tool for building accurate CG models of biomolecules. Since the Martini~3 lipidome
provides a systematic classification and mapping of lipid chemistry, developing a transferable
MLCG lipid model at Martini~3 resolution is a natural and achievable next step. Moreover, because
real membranes contain proteins and other molecules in addition to lipids, coupling such a lipid
model to a transferable MLCG protein model would open the door to simulating realistic membranes.
Such a lipid--protein model would be a step toward a foundation model for biomolecules, enabling
scalable simulations of biomolecular complexes and providing insight into the molecular mechanisms
underlying biological function.

\section*{Data Availability}
The core library used in this work is available on GitHub at \url{https://github.com/ClementiGroup/mlcg}.
The datasets and models will be released upon publication.
\section*{Acknowledgments}
A.S.P.T. and C.C. acknowledge support from the Deutsche Forschungsgemeinschaft DFG 
(SFB/TRR 186, project A12; SFB 1114, projects B03, B08, and A04), the Einstein Foundation Berlin (project 0420815101), the Bundesministerium
f\"ur Bildung und Forschung BMBF (project FAIME 01IS24076) and from the International Max Planck Research School for
Biology and Computation (IMPRS-BAC). D.P.T. acknowledges support from the
Natural Sciences and Engineering Research Council of Canada (NSERC) and the Canada Research Chairs program. D.C.T.
acknowledges support from the Spanish Ministerio de Universidades through a
predoctoral contract (FPU22/00636). The computations in this work were performed
using resources of the Digital Research Alliance of Canada, supported by the
Canada Foundation for Innovation and partners (D.P.T.), and on the GCS
supercomputers JUWELS~\cite{JUWELS} and JUPITER at the Jülich Supercomputing
Centre (JSC), provided by the Gauss Centre for Supercomputing e.V.
(www.gauss-centre.eu).
\clearpage
\setcounter{secnumdepth}{3}
\setcounter{section}{0}
\setcounter{figure}{0}
\setcounter{table}{0}
\setcounter{equation}{0}
\renewcommand{\thesection}{S\arabic{section}}
\renewcommand{\thesubsection}{\thesection.\arabic{subsection}}
\renewcommand{\thefigure}{S\arabic{figure}}
\renewcommand{\thetable}{S\arabic{table}}
\renewcommand{\theequation}{S\arabic{equation}}
\setcounter{page}{1}
\renewcommand{\thepage}{S\arabic{page}}

\begin{center}
  {\LARGE\bfseries Supporting Information\par}
  \vspace{1em}
  {\Large \papertitle\par}
  \vspace{1em}
  {\large Aldo S. Pasos-Trejo, Daniel P. Ramirez-Echemendia, Daniel Conde-Torres,
   D. Peter Tieleman, Cecilia Clementi*\par}
\end{center}
\vspace{2em}

\section{Methods details}
\subsection{Dataset and mapping}\label{sec:si-dataset}
Atomistic reference trajectories of POPC were generated with the AMBER Lipid17
FF~\cite{gouldLipid17ComprehensiveAMBER} and the OpenMM MD
engine~\cite{eastmanOpenMM8Molecular2024}. Initial bilayer configurations were
assembled with the CHARMM-GUI Membrane
Builder~\cite{leeCHARMMGUIMembraneBuilder2019} and solvated with 3~nm of TIP3P
water molecules on both sides of the membrane. The single-lipid system was
extracted from the equilibrated bilayer and re-solvated with the GROMACS
\texttt{solvate} tool~\cite{abrahamGROMACSHighPerformance2015} to fill a cubic
$5 \times 5 \times 5$~nm$^{3}$ box. All systems
were energy-minimized, equilibrated, and propagated at $T = 300$~K in the NVT
ensemble using a Langevin-middle
integrator~\cite{zhangUnifiedEfficientThermostat2019} with a friction
coefficient of 1~ps$^{-1}$ and a 4~fs timestep, enabled by hydrogen mass
repartitioning~\cite{hopkinsLongTimeStepMolecularDynamics2015} with a hydrogen
mass of 4~amu. Non-bonded
interactions used a 0.9~nm cutoff, long-range electrostatics were treated with
particle-mesh Ewald~\cite{dardenParticleMeshEwald1993}, and bonds involving
hydrogens were constrained. 

The Martini~3 lipidome representation~\cite{pedersenMartini3Lipidome2025} maps
disjoint sets of atoms onto CG beads, using 12 beads per lipid molecule.
Table~\ref{tab:popc-mapping} lists the atoms assigned to each CG bead. For every
atomistic configuration, each CG bead is placed at the center of geometry of its
assigned atoms. CG target forces were obtained with a separate, optimal linear
force map $\mathcal{M}_F$ constructed with the \texttt{aggforce}
library~\cite{kramer_statistically_2023}. This map is the matrix of minimum
Frobenius norm that satisfies the Ciccotti--Hansen consistency constraint with
$\mathcal{M}_C$. It therefore yields mapped forces
$\mathbf{F}_{\text{CG}} = \mathcal{M}_F\,\mathbf{f}_{\text{AA}}$ whose Boltzmann
average still recovers the gradient of the exact many-body PMF, while minimizing
the variance of the force-matching residual~\cite{kramer_statistically_2023}.
Water and ions are not represented by any CG site, so their contribution to the
PMF is absorbed implicitly through the optimized force projection.
\begin{table}[H]
\centering
\caption{Mapping of POPC atoms onto Martini~3 CG beads.}
\label{tab:popc-mapping}
\begin{tabular}{cp{14cm}}
\hline
CG bead & Atoms \\
\hline
NC3 & N, C13, H13A, H13B, H13C, C14, H14A, H14B, H14C, C15, H15A, H15B, H15C, C12, H12A, H12B, C11, H11A, H11B \\
PO4 & P, O13, O14, O11, O12 \\
GL1 & C2, HS, O21, C21, O22 \\
GL2 & C3, HX, HY, O31, O32, C31 \\
C1A & C22, H2R, H2S, C23, H3R, H3S, C24, H4R, H4S, C25, H5R, H5S, C26, H6R, H6S \\
D2A & C27, H7R, H7S, C28, H8R, H8S, C29, H91, C210, H101 \\
C3A & C211, H11R, H11S, C212, H12R, H12S, C213, H13R, H13S, C214, H14R, H14S \\
C4A & C215, H15R, H15S, C216, H16R, H16S, C217, H17R, H17S, C218, H18R, H18S, H18T \\
C1B & C32, H2Y, H2X, C33, H3Y, H3X, C34, H4Y, H4X \\
C2B & C35, H5Y, H5X, C36, H6Y, H6X, C37, H7Y, H7X, C38, H8Y, H8X \\
C3B & C39, H9X, H9Y, C310, H10X, H10Y, C311, H11Y, H11X, C312, H12Y, H12X \\
C4B & C313, H13Y, H13X, C314, H14Y, H14X, C315, H15Y, H15X, C316, H16Y, H16X, H16Z \\
\hline
\end{tabular}
\end{table}
\newpage
\subsection{Prior fitting} \label{si:prior-fitting}
\subsubsection{Bond prior}
For every bonded pair $(i,j)$ of beads, the bond prior is a harmonic potential
in the bond length $r_{ij}$, with parameters that depend on the bead types
$(a,b)$:
\begin{equation}
  U_{\mathrm{bond}}(r_{ij}) = k_{ab}\left(r_{ij} - r_{0,ab}\right)^2 .
  \label{eq:harmonic_bond}
\end{equation}
For each bond type, we built a histogram $h_{ab}(r)$ of bond lengths in the
CG-mapped atomistic reference trajectory, with 500 uniform bins between 2 and
6~\AA. Boltzmann inversion gave a potential of mean force, $V_{ab}(r) =
-k_\mathrm{B}T\ln h_{ab}(r)$, at $T=300$~K. We discarded empty bins and bins
with $|V_{ab}| < 10^{-4}\left|\int V_{ab}\,\mathrm{d}r\right|$. The parameters
$k_{ab}$ and $r_{0,ab}$ were obtained by non-linear least squares
(concretely, \texttt{scipy.optimize.curve\_fit} from the SciPy package~\cite{2020SciPy}), fitting $k_{ab}(r-r_{0,ab})^2 + V_{0,ab}$
to $V_{ab}(r)$ with the fit initialized at the minimum of $V_{ab}$. The constant
$V_{0,ab}$ was discarded as it doesn't affect the force computation.

\subsubsection{Angle prior}
For every bonded triplet $(i,j,k)$ of beads, the angle prior is a fourth-order
polynomial in the bond angle $\theta_{ijk}$ (in radians), with coefficients that
depend on the bead types $(a,b,c)$:
\begin{equation}
  U_{\mathrm{ang}}(\theta_{ijk}) = V_{0,abc} + \sum_{n=1}^{4} k_{n,abc}\,\theta_{ijk}^{\,n}.
  \label{eq:quartic_angle}
\end{equation}
For each angle type, we built a normalized histogram $p_{abc}(\theta)$ of bond
angles in the CG-mapped atomistic reference trajectory, with 41 uniform bins
over the sampled range. Analogously to the bond prior, we obtained the potential
of mean force by Boltzmann inversion and discard the empty bins following the integral criterion mentioned. 
Because Eq.~\eqref{eq:quartic_angle} is
linear in its coefficients, $V_{0,abc}$ and $k_{n,abc}$ were obtained directly
by linear least squares at the bin centers.

\subsubsection{Non-bonded prior.}
The non-bonded prior acts on all bead pairs $(i,j)$ in the non-bonded neighbor
list. These are the pairs not already covered by the bonded and angle priors. It
is a pairwise exponential repulsion whose parameters depend only on the bead
types $a(i)$ and $b(j)$:
\begin{equation}
  U_{\mathrm{nb}}(r_{ij}) = \frac{6}{\alpha_{ab}}
  \exp\!\left[\alpha_{ab}\left(1-\frac{r_{ij}}{r_{0,ab}}\right)\right].
  \label{eq:exp_repulsion}
\end{equation}
For each of the 78 unique pairs of the 12 bead types, we fitted $\alpha_{ab}$
and $r_{0,ab}$ as follows. First, we built a histogram $h_{ab}(r)$ of pair
distances from the CG-mapped reference trajectory, with 500 uniform bins between
0 and 10~\AA. For the non-bonded interaction, we only consider the pairs that are at least 4 beads 
apart in the bond-graph of the system (i.e. not constrained by a bond or angle term). 
Histograms of the ordered pairs $(a,b)$ and $(b,a)$ were summed.
We obtained a potential of mean force by Boltzmann inversion, $V_{ab}(r) =
-k_\mathrm{B}T\ln h_{ab}(r)$, using only non-empty bins. We
shifted it so that it vanishes at the largest sampled distance and smoothed it
with a Savitzky--Golay filter (window of 5 bins, third-order polynomial).
Eq.~\eqref{eq:exp_repulsion} describes only the short-range repulsive wall, so
the fit was restricted to distances below the first structural feature of
$V_{ab}$. To find this point, we fitted a fourth-order polynomial to $V_{ab}(r)$
for $r<8$~\AA. We then took the smallest local minimum or inflection point of
this polynomial within the sampled range, $r^{*}_{ab}$. We fitted $\alpha_{ab}$
and $r_{0,ab}$ to the smoothed $V_{ab}(r)$ on $r<r^{*}_{ab}$ with a genetic
algorithm (concretely, \texttt{pygad.GA} from the PyGAD library~\cite{gad2023pygad}), 
using the inverse $\ell_2$ norm of the residual as the fitness. The algorithm ran 
for 200 generations with a population of 80, initial parameters drawn uniformly 
from $[1,10]$, steady-state selection of 20 parents,
arithmetic-mean crossover and random mutation of 50\% of the genes. As a repulsion
fitted directly through this method remained too strong to be used in simulation,
we proceed to scale down the parameters by a factor of 0.72 and 0.9 for $r_{0,ab}$ 
and $\alpha_{ab}$, respectively.
Finally, the potential was truncated at the GNN cutoff $r_c = 12$~\AA{} with a shifted-force
scheme, so that both the energy and the force go continuously to zero at the
cutoff:
\begin{equation}
  U^{c}_{\mathrm{nb}}(r) =
  \begin{cases}
    U_{\mathrm{nb}}(r) - U_{\mathrm{nb}}(r_c) - (r-r_c)\,U'_{\mathrm{nb}}(r_c), & r < r_c,\\
    0, & r \geq r_c.
  \end{cases}
\end{equation}
\subsection{Model hyperparameters and training}~\label{si:hyperparams}
Both models were implemented and trained with the \texttt{mlcg} package and PyTorch
Lightning. Table~\ref{tab:hyperparams} lists the architecture and training
hyperparameters. The two models differ in the number of interaction blocks and in
the composition of the minibatches. For \texttt{Lipid-MLCG}, each minibatch combines
frames from both training systems (8 bilayer frames and 160 single-lipid frames per
GPU, see Figure~\ref{fig:dataset-ablation} for other compositions). For both
models, the data were split into training and validation sets in an 80/20 ratio.
\begin{table}[H]
\centering
\caption{Hyperparameters of the SchNet architecture and training, following the notation of Ref.~\cite{pasostrejo2026exploration}.}
\label{tab:hyperparams}
\begin{tabular}{p{5cm}cc}
\hline
Hyperparameter & \texttt{SinLip-MLCG} & \texttt{Lipid-MLCG} \\
\hline
 Embedding feature size ($H$) & 128 & 128 \\
 Number of basis functions ($M$) & 64 & 64 \\
 Number of filters ($L$) & 128 & 128 \\
 Interaction blocks ($T$) & 2 & 3 \\
 Distance cutoff ($r_c$) & 12~\AA{} & 12~\AA{} \\
 Cutoff function & Cosine & Cosine \\
 Radial basis function & PhysNet~\cite{Unke_Meuwly_2019} & PhysNet~\cite{Unke_Meuwly_2019} \\
 Activation function & tanh & tanh \\
 EnerMLP layers & [128, 64] & [128, 64] \\
\hline
 Optimizer & Adam & Adam \\
 Learning rate & $10^{-4}$ & $10^{-4}$ \\
 Weight decay & 0 & 0 \\
 Numerical precision & FP32 & FP32 \\
 GPUs & 1 & 4 (data parallel) \\
 Batch size per GPU & 256 & 8 bilayer + 160 single-lipid frames \\
 Training/validation split & 80/20 & 80/20 \\
\hline
\end{tabular}
\end{table}
\subsection{Simulation details}
For the CG Langevin simulations, we used a BAOAB integrator~\cite{leimkuhler2015molecular}
with a timestep of 16~fs and a friction coefficient of 1~ps$^{-1}$.All CG simulations,
including the DOPC, DPPC and DOPC/DPPC systems, were run at 300~K, the temperature of the
atomistic training data. CG bead masses are the
same as in Martini~3. Periodic boundary conditions were enforced using the minimum image
convention. Simulation times of the \texttt{Lipid-MLCG} model are given in CG time, the number of
integration steps times the 16~fs timestep, and written cg-ns. Because the CG dynamics are
accelerated (Table~\ref{tab:bilayer-metrics}), CG time does not correspond to physical time.
\par
\begin{figure}
    \centering
    \includegraphics[width=\linewidth]{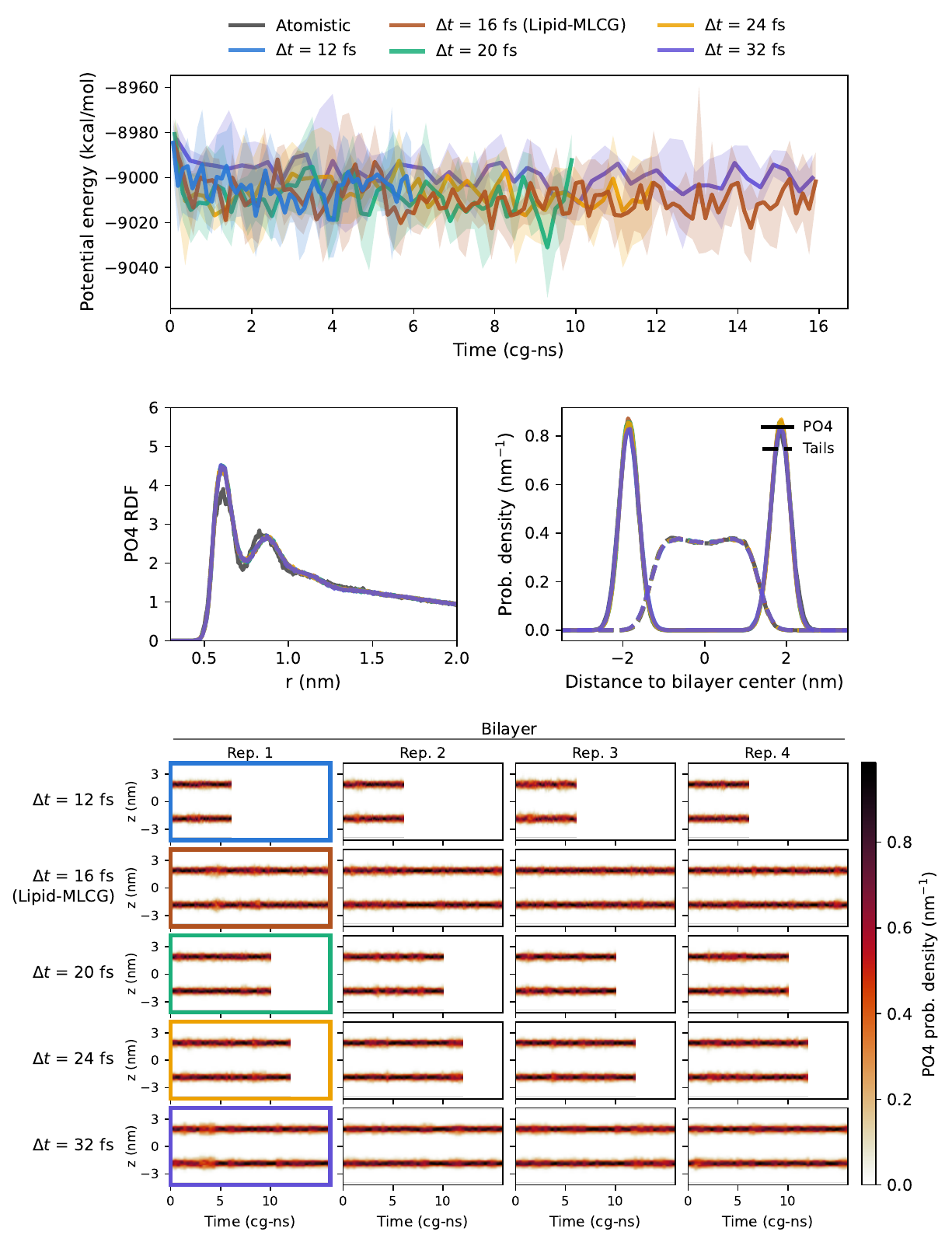}
    \caption{Sensitivity to the integration timestep: \texttt{Lipid-MLCG} bilayer
simulations with timesteps $\Delta t$ of 12, 16 (\texttt{Lipid-MLCG}),
20, 24 and 32~fs. The runs with modified timesteps have $5\times10^5$ integration steps, so they
cover 6 to 16~cg-ns. (Top) Total potential energy as a function of time, averaged over windows of
10 saved frames. Lines show the mean over runs and shaded bands the range between runs.
(Middle) headgroup radial distribution function and headgroup (solid) and tail-bead (dashed) density profiles
along the bilayer normal, combining all replicas, compared with the atomistic reference (gray).
(Bottom) headgroup density along $z$ as a function of time for each replica.}
    \label{fig:dt-ablation}
\end{figure}

It should be noted that the timestep of 16~fs is selected for stability but it could in principle be
increased to at least double without losing the stable bilayer (Figure~\ref{fig:dt-ablation})
\subsection{Martini reference simulations}
The Martini~3~\cite{pedersenMartini3Lipidome2025} and DryMartini~\cite{arnarez2015dry} reference
simulations used the same 150-lipid POPC bilayer as the atomistic reference and were run with GROMACS~\cite{abrahamGROMACSHighPerformance2015} for 1~\textmu s
in the NVT ensemble at 303.15~K, using a 20~fs timestep and reaction-field electrostatics
($\epsilon_r = 15$). For Martini~3, we used the leap-frog integrator with the velocity-rescaling
thermostat~\cite{bussiCanonicalSamplingVelocity2007} ($\tau_T = 1$~ps) and a 1.1~nm cutoff for
both Coulomb and Lennard-Jones interactions. For DryMartini, we used the stochastic dynamics integrator
($\tau_T = 4$~ps) and a 1.2~nm cutoff.

\subsection{Analysis of bilayer structure and dynamics}\label{sec:si-bilayer-analysis}

All quantities in Table~\ref{tab:bilayer-metrics} and Figure~\ref{fig:bilayer} were
computed on the CG representation. The atomistic trajectory was first mapped to the
Martini~3 resolution (each bead at the center of geometry of its atoms, see
Section~\ref{sec:si-dataset}), so that all models are analyzed with exactly the same
bead definitions. The analysis was carried out with
MDAnalysis~\cite{gowersMDAnalysisPythonPackage2016} and
LiPyphilic~\cite{smithLiPyphilicPythonToolkit2021}.

\subsubsection{Leaflet assignment and bilayer thickness}
In every frame, each lipid was assigned to the upper or lower leaflet according to
the position of its PO4 bead along $z$ relative to the bilayer midplane, defined as
the mean $z$ of all PO4 beads.

The headgroup--headgroup thickness was computed from the PO4 beads. In each frame,
the beads were wrapped into the primary simulation box, and each lipid was assigned
to the upper ($\mathcal{U}$) or lower ($\mathcal{L}$) leaflet depending on whether
its PO4 bead lies above or below the bilayer midplane,
$\bar{z}(t) = N^{-1}\sum_{i=1}^{N} z_{\text{PO4},i}(t)$, the mean $z$ of all $N$
PO4 beads. The thickness is the distance between the mean heights of the two
leaflets,
\begin{equation}
  d_{HH}(t) = \frac{1}{N_{\mathcal{U}}} \sum_{i\in\mathcal{U}} z_{\text{PO4},i}(t)
            - \frac{1}{N_{\mathcal{L}}} \sum_{i\in\mathcal{L}} z_{\text{PO4},i}(t),
  \label{eq:si-dhh}
\end{equation}
where $N_{\mathcal{U}}$ and $N_{\mathcal{L}}$ are the numbers of lipids in each leaflet.
Because a single global height is used for each leaflet, this definition assumes a
flat bilayer perpendicular to $z$ that does not cross the periodic boundary along
$z$, which holds for all the flat-bilayer simulations analyzed here.

\subsubsection{Tail order parameter}

The tail order parameter was computed for both acyl chains,
\begin{equation}
  S_{CC}(t) = \left\langle \frac{3\cos^2\theta_{k}(t) - 1}{2} \right\rangle,
  \label{eq:si-scc}
\end{equation}
where $\theta_k$ is the angle between the bond $k$ joining two consecutive beads of a
chain and the $z$ axis (the bilayer normal), and the average runs over all bonds and
all lipids. The chains were defined as GL1--C1A--D2A--C3A--C4A (\textit{sn}-2,
oleoyl) and GL2--C1B--C2B--C3B--C4B (\textit{sn}-1, palmitoyl), and the reported
$S_{CC}$ is the mean of the two chains. Values in Table~\ref{tab:bilayer-metrics} are
the mean and standard deviation over the frames of the first 80~ns of each trajectory.

\subsubsection{Lateral diffusion.}
Lateral diffusion was characterized from the in-plane ($xy$) positions of the PO4
beads. Trajectories were unwrapped by accumulating the minimum-image displacement
between consecutive frames, and in every frame the center-of-mass position of each
leaflet was subtracted from its lipids to remove the collective drift of the
leaflets. For each lag time $\tau$, we collected the lateral displacements
$r = |\Delta\mathbf{r}_{xy}(\tau)|$ of all lipids over all available time origins,
which gives the displacement distributions $P(r,\tau)$ shown in
Figure~\ref{fig:bilayer}B, and their mean squared value
$\langle r^2(\tau) \rangle$. The same set of absolute lag times,
$\tau = 8, 12, 16, 20$ and $24$~ns, was used for all models. Because lateral motion is
subdiffusive at short times, an effective diffusion coefficient depends on the lag
times at which it is measured, so fixing $\tau$ is necessary for a meaningful
comparison between models. The diffusion coefficient was obtained from a linear fit
\begin{equation}
  \langle r^2(\tau) \rangle = 4 D \tau + b,
  \label{eq:si-msd}
\end{equation}
with a free intercept $b$, which absorbs the short-time subdiffusive contribution
that would otherwise bias the slope; the uncertainty of $D$ is the standard error of
the fitted slope divided by 4. To quantify how far each model is from the Fickian
regime within this window, we also fitted the anomalous exponent $\alpha$ in
$\langle r^2(\tau) \rangle \propto \tau^{\alpha}$ by linear regression of
$\ln\langle r^2 \rangle$ against $\ln\tau$ over the same lag times; $\alpha = 1$
corresponds to normal diffusion and $\alpha < 1$ to subdiffusion. For comparison with
the measured distributions, Figure~\ref{fig:bilayer}B shows the two-dimensional
Gaussian (Rayleigh) distribution
$P(r,\tau) = (r/\sigma^2)\exp\!\left[-r^2/(2\sigma^2)\right]$ with
$2\sigma^2 = 4D\tau + b$, i.e., the distribution expected for normal diffusion with
the fitted parameters. The full length of each trajectory was used to average over
time origins.

\subsection{Analysis of curved systems}\label{sec:si-curved-analysis}
For the vesicles and the micelle (Tables~\ref{tab:curved-systems},
\ref{tab:curved-metrics} and~\ref{tab:complex-metrics}), all quantities were computed in a radial
frame centered on the aggregate: the center of mass of all beads for the preformed systems,
which were simulated without periodic boundaries, and the periodic (circular) mean position of
the vesicle beads for the self-assembled vesicles, whose positions were then unwrapped with the
minimum-image convention. A lipid was assigned to the inner leaflet when its PO4 bead is closer
to the center than the midpoint of its two tail ends (C4A, C4B), i.e., when its headgroup points
inwards. This definition requires no radius threshold and follows a vesicle that changes shape. The leaflet
radii $R_\text{out}$ and $R_\text{in}$ are the median PO4 distances to the center in each
leaflet, which are robust to ejected lipids, and $d_{HH} = R_\text{out} - R_\text{in}$; for
the micelle, the radius is the mean PO4 distance. The area per lipid of each leaflet is
$4\pi R^2/N$, with $N$ the number of lipids in that leaflet. $S_{CC}$ is
Eq.~\eqref{eq:si-scc} with $\theta_k$ the angle between bond $k$ and the radial direction at the
bond midpoint, and $P_2$ is Eq.~\eqref{eq:si-p2} with the radial direction at the PO4 bead. The
shape ratio $a_\text{max}/a_\text{min} = \sqrt{\lambda_\text{max}/\lambda_\text{min}}$ is
computed from the largest and smallest principal moments of the gyration tensor of the PO4 beads, 
and should equal 1 for a perfect sphere.

For the self-assembled vesicles, lipids were grouped into aggregates (two lipids belong to the
same aggregate when any two of their tail beads are closer than 0.7~nm, with periodic
boundaries), and the vesicle was taken as the largest aggregate; the fraction of lipids in the
vesicle given in the main text is the fraction in the largest aggregate. The vesicle metrics
were computed over the last 10\% of each run, using only the lipids that belong to the largest
aggregate in every frame of that window. Table~\ref{tab:vesicle-self-assembly} compares in detail
the size, shape and leaflet composition of the self-assembled vesicles with those of the preformed small vesicle.

\begin{table}[htbp]

  \centering
  \caption{Size, shape and leaflet composition of the vesicles obtained by self-assembly with
    \texttt{Lipid-MLCG} (8 runs), compared with the preformed small vesicle with the same number of
    lipids (959). $R_\text{out}$, $R_\text{in}$ and $a_\text{max}/a_\text{min}$ are defined as in
    Table~\ref{tab:curved-metrics}. $N_\text{out}/N_\text{in}$ is the number of lipids in the outer over
    the inner leaflet, which is fixed by construction in the preformed vesicle and is an outcome of the
    assembly in the self-assembled ones. The last column gives the percentage of lipids in the largest
    aggregate (the vesicle). Values are the mean $\pm$ standard deviation over frames, pooled over runs, over the same frames as
    Table~\ref{tab:complex-metrics}.}
  \label{tab:vesicle-self-assembly}
  \begin{tabular}{lccccc}
    \hline
    System & $R_\text{out}$ (nm) & $R_\text{in}$ (nm) & $a_\text{max}/a_\text{min}$ & $N_\text{out}/N_\text{in}$ & Lipids in vesicle (\%) \\
    \hline
    Preformed small vesicle & $7.15 \pm 0.02$ & $3.91 \pm 0.03$ & $1.08 \pm 0.03$ & 2.10 & 100 \\
    Self-assembled vesicle  & $7.03 \pm 0.04$ & $3.76 \pm 0.05$ & $1.08 \pm 0.04$ & $2.44 \pm 0.07$ & $97 \pm 2$ \\
    \hline
  \end{tabular}
\end{table}

To locate the disorder induced by curvature, Table~\ref{tab:curved-leaflets} splits the
structure of the preformed vesicles by leaflet. The leaflet thickness $h$ is the distance along the
radial direction between the median PO4 radius and the median radius of the tail-end midpoints of
that leaflet; $d_{HH}$ equals the sum of the two leaflet thicknesses and the gap between the tail
ends of the two leaflets (5.3~\AA{} in the flat bilayer, 4.8~\AA{} in both vesicles). $S_{CC}$ and $P_2$ are
given with respect to the radial direction, as in Table~\ref{tab:curved-metrics}, and with respect to
a local normal: the mean director, oriented outwards, of all lipids whose PO4 bead lies within 2~nm
of that of the lipid, measured along the vesicle surface (the curved counterpart of the local normal
used for the sealed pores, Section~\ref{sec:si-selfassembly-analysis}). With either reference
direction, both leaflets of both vesicles remain less ordered than the flat bilayer, so the reduction
of $S_{CC}$ reflects chain disorder rather than the choice of reference direction.
The disorder is concentrated in the inner leaflet, which is the thinnest and least ordered leaflet
in both vesicles. Because the inner leaflet is concave, its headgroups are closely packed while its tails extend
into a widening volume: the area per lipid at the tail ends is 40--65\% larger than in the flat
bilayer. In the outer leaflet the situation is reversed, and the area per lipid at the tail ends
is close to that of the flat bilayer, so its chains remain more ordered despite the larger area per
lipid at the PO4 surface.

\begin{table}[htbp]
  \centering
  \caption{Per-leaflet structure of the preformed vesicles simulated with \texttt{Lipid-MLCG},
    compared with the flat bilayer (both leaflets pooled). $h$ is the leaflet thickness
    (median PO4 minus median tail-end-midpoint coordinate along the normal). $S_{CC}$ and $P_2$ are
    computed with respect to the radial direction (flat bilayer: $z$), as in
    Table~\ref{tab:curved-metrics}, and $S_{CC}^\text{loc}$ and $P_2^\text{loc}$ with respect to a
    local normal (see text). The area per lipid is $4\pi R^2/N$ at the median PO4 radius
    (APL$_\text{PO4}$, as in Table~\ref{tab:curved-metrics}) and at the median radius of the tail-end
    midpoints (APL$_\text{end}$) of each leaflet; for the flat bilayer, both are the box area per
    leaflet lipid. Values are the mean $\pm$ standard deviation over frames, excluding the first
    1~cg-ns, pooled over 4 independent runs. The standard deviation of the APL is omitted as its 
    at most 0.011~nm$^2$.}
  \label{tab:curved-leaflets}
  \small
  \setlength{\tabcolsep}{4pt}
  \begin{tabular}{llccccccc}
    \hline
    System & Leaflet & $h$ (\AA{}) & $S_{CC}$ & $S_{CC}^\text{loc}$ & $P_2$ & $P_2^\text{loc}$ & APL$_\text{PO4}$ (nm$^2$) & APL$_\text{end}$ (nm$^2$) \\
    \hline
    Flat bilayer  & both  & $15.9 \pm 0.4$ & $0.349 \pm 0.017$ & $0.355 \pm 0.016$ & $0.79 \pm 0.02$ & $0.80 \pm 0.02$ & 0.67 & 0.67 \\
    Large vesicle & inner & $13.7 \pm 0.1$ & $0.246 \pm 0.005$ & $0.227 \pm 0.005$ & $0.67 \pm 0.01$ & $0.70 \pm 0.01$ & 0.68 & 0.95 \\
                  & outer & $14.7 \pm 0.1$ & $0.281 \pm 0.004$ & $0.309 \pm 0.004$ & $0.76 \pm 0.01$ & $0.77 \pm 0.00$ & 0.89 & 0.67 \\
    Small vesicle & inner & $13.1 \pm 0.2$ & $0.229 \pm 0.010$ & $0.188 \pm 0.010$ & $0.62 \pm 0.02$ & $0.64 \pm 0.01$ & 0.62 & 1.11 \\
                  & outer & $14.4 \pm 0.2$ & $0.268 \pm 0.006$ & $0.307 \pm 0.006$ & $0.75 \pm 0.01$ & $0.77 \pm 0.01$ & 0.99 & 0.63 \\
    \hline
  \end{tabular}
\end{table}

\subsection{Structural analysis of self-assembly}\label{sec:si-selfassembly-analysis}

\subsubsection{Bilayer normal and tail structure factor.}
Bilayers formed by self-assembly in a periodic box need not be perpendicular to $z$, as
they can form along any lattice direction of the box. To identify the bilayer normal
without assuming its orientation, we computed the structure factor of the acyl-chain
beads (C1A, D2A, C3A, C4A, C1B, C2B, C3B, C4B),
\begin{equation}
  S(\mathbf{q}, t) = \frac{1}{N_\text{T}} \sum_{j=1}^{N_\text{T}}
  \exp\!\left[\,i\,\mathbf{q}\cdot\mathbf{r}_j(t)\right],
  \label{eq:si-sq}
\end{equation}
where the sum runs over all $N_\text{T}$ tail beads and $\mathbf{r}_j(t)$ is the
position of bead $j$ at time $t$. The wave vectors are restricted to the lowest
harmonics compatible with the orthorhombic box of edge lengths $L_x, L_y, L_z$,
$\mathbf{q}_{hkl} = 2\pi\,(h/L_x,\, k/L_y,\, l/L_z)$ with $h,k,l \in \{-1,0,1\}$,
which gives 13 distinct directions once $\mathbf{q}$ and $-\mathbf{q}$ are identified.
Because the tails of a bilayer form one slab per periodic repeat, $|S(\mathbf{q},t)|$
approaches 1 when the tails are confined to a thin layer perpendicular to
$\mathbf{q}$, and vanishes for a homogeneous (dispersed) distribution. The bilayer
normal $\hat{\mathbf{n}}$ of each trajectory was taken as the direction
$\mathbf{q}_{hkl}$ that maximizes $|S|$, averaged over the last 10\% of the frames.
Since the tails occupy a slab of finite thickness, $|S| < 1$ even for a perfect
bilayer; in our simulations, values of $|S| \gtrsim 0.65$ correspond to a well-formed
bilayer, whereas $|S| \approx 0.4$--$0.5$ indicates two broad, weakly separated layers.
For diagonal directions the repeat distance $d = 2\pi/|\mathbf{q}|$ is shorter and the
slab fills a larger fraction of it, so $|S|$ values are not strictly comparable
between directions.

\subsubsection{Density and thickness along the bilayer normal.}\label{sec:density-along-bilayer-normal}
The PO4 density maps were computed along the selected normal. For each bead we
computed the phase $\phi_j = \mathbf{q}\cdot\mathbf{r}_j$ relative to the circular
(periodic) mean phase of all beads, $\bar{\phi} = \arg \sum_j e^{i\phi_j}$, wrapped it
to $(-\pi, \pi]$, and converted it to a distance
$h_j = (\phi_j - \bar{\phi})/|\mathbf{q}|$, which lies in $[-d/2, d/2]$. For an intact
bilayer along $z$, this equals the $z$ coordinate relative to the center of mass,
but, unlike the latter, it does not split a bilayer that crosses the periodic boundary.
For the self-assembled and sealed bilayers (Table~\ref{tab:complex-metrics}), $d_{HH}$ is
computed as in Eq.~\eqref{eq:si-dhh}, with $z_{\text{PO4},i}$ replaced by $h_{\text{PO4},i}$, the
value of $h_j$ at the PO4 bead of lipid $i$: each lipid is assigned to the upper ($\mathcal{U}$) or
lower ($\mathcal{L}$) leaflet depending on whether $h_{\text{PO4},i}$ lies above or below the mean
$\bar{h} = N^{-1}\sum_{i=1}^{N} h_{\text{PO4},i}$ of all $N$ PO4 beads, and $d_{HH}$ is the
difference between the mean $h_{\text{PO4},i}$ of the two leaflets.

\subsubsection{Lipid orientational order $P_2$.}
To follow how lipids align during assembly, we defined for each lipid $i$ a director
from its phosphate bead to the midpoint of its two tail ends,
\begin{equation}
  \mathbf{u}_i = \tfrac{1}{2}\left[(\mathbf{r}_{\text{C4A},i} - \mathbf{r}_{\text{PO4},i})
  + (\mathbf{r}_{\text{C4B},i} - \mathbf{r}_{\text{PO4},i})\right],
\end{equation}
using the minimum-image convention, and computed the orientational order parameter
\begin{equation}
  P_2(t) = \frac{1}{N} \sum_{i=1}^{N} \frac{3\cos^2\theta_i(t) - 1}{2},
  \qquad
  \cos\theta_i = \frac{\mathbf{u}_i \cdot \hat{\mathbf{n}}_i}{|\mathbf{u}_i|},
  \label{eq:si-p2}
\end{equation}
where $\hat{\mathbf{n}}_i$ is the reference direction for lipid $i$. For planar bilayers,
$\hat{\mathbf{n}}_i$ is the bilayer normal $\mathbf{q}_{hkl}/|\mathbf{q}_{hkl}|$ defined above;
for vesicles, it is the radial direction from the center of the largest aggregate to the
PO4 bead of lipid $i$, and the average runs over the lipids of that aggregate only
(Section~\ref{sec:si-curved-analysis}. For bicelles, it is the principal axis of smallest variance of
the largest aggregate (lipids are assigned to the same aggregate when any two of their
tail beads are closer than 0.7~nm), and the average runs over the lipids of that
aggregate only. For the monolayer simulations, $\hat{\mathbf{n}}_i = \hat{\mathbf{z}}$.
$P_2 = 1$ for lipids perfectly aligned with the reference direction, $P_2 = 0$ for an
isotropic distribution, and $P_2 = -1/2$ for lipids perpendicular to it. Because
$P_2$ is even in $\cos\theta_i$, it does not distinguish lipids pointing up from lipids
pointing down. In systems such as the monolayer simulations, where this pointing direction 
matters, we report the fraction of lipids with their headgroup pointing towards $+z$ separately.

It should be noted that $P_2$ differs from the tail order parameter $S_{CC}$ reported in the main text. 
$S_{CC}$ averages the same Legendre polynomial over the individual bonds between consecutive
tail beads and therefore measures the conformational order of the acyl chains.
$P_2$ uses a single vector per lipid and measures the orientation of the molecule as a
whole. A lipid can be well aligned with the normal ($P_2 \approx 1$) while its chains
remain conformationally disordered (low $S_{CC}$). Accordingly, the intact bilayer
has $P_2 = 0.79$ but $S_{CC} = 0.35$ (Table~\ref{tab:bilayer-metrics}), and the self-assembled
vesicles $P_2 = 0.71$ but $S_{CC} = 0.27$ (Table~\ref{tab:complex-metrics}).

\subsubsection{Area per lipid, assembly time and pore sealing.}\label{sec:apl-assembly-time-and-pore-sealing}
Because the tails of a planar bilayer form one slab per periodic repeat
$d = 2\pi/|\mathbf{q}_{hkl}|$, a bilayer along $\mathbf{q}_{hkl}$ covers an area $V/d$ of a box
of volume $V$, and its area per lipid is $\text{APL} = V/(d\,N/2)$; for $(hkl) = (001)$ this
reduces to $L_x L_y/(N/2)$.

\begin{figure}
    \centering
    \includegraphics[width=0.9\linewidth]{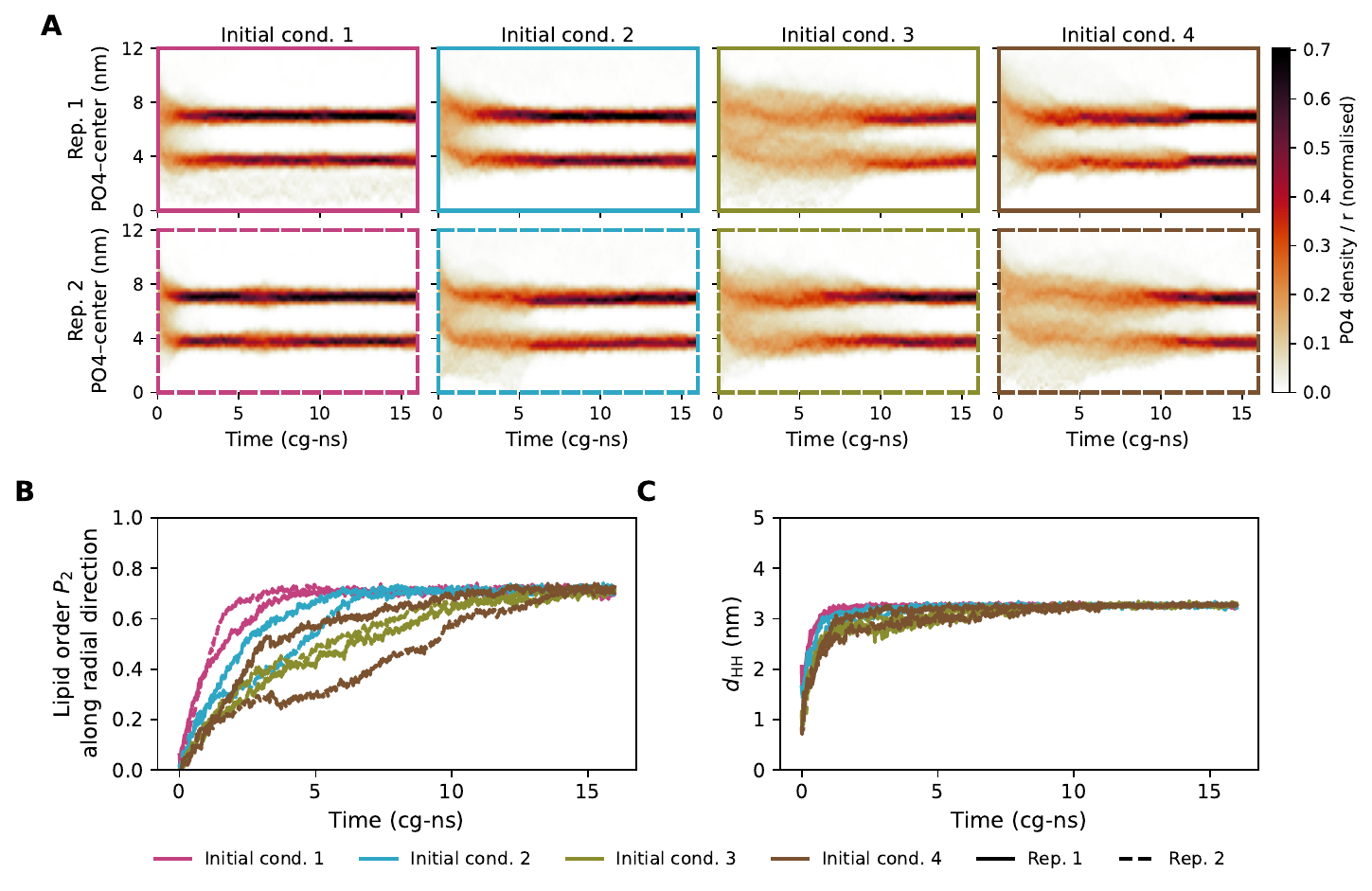}
    \caption{Vesicle self-assembly from the medium-density, spherically distributed system.
    (A) Heatmap of the headgroup density as function of the simulation time for 4 different initial conditions 
    with 2 replicas each. 
    (B) Lipid ordering parameters $P_2$ along the aggregate normal and (C) $d_{HH}$ the bilayer thickness. 
    Following Table~\ref{tab:complex-metrics}, both $P_2$ and $d_{HH}$ are computed in every 
    frame for the lipids of the largest aggregate, see Section~\ref{sec:si-curved-analysis}.}
    \label{fig:vesicle_self_assembly}
\end{figure}

We define he assembly time as the time after which the lipids stay aligned. We smoothed $P_2(t)$
(Eq.~\eqref{eq:si-p2}; for vesicles, over the lipids of the largest aggregate, as plotted in
Figure~\ref{fig:vesicle_self_assembly}) with a running mean over 25 frames (0.4~cg-ns) and took the assembly time as the first
time after which the smoothed $P_2$ stays above $P_2^0 + 0.9\,(P_2^\infty - P_2^0)$, where $P_2^0$
is its mean over the first 25 frames and $P_2^\infty$ the mean of $P_2$ over the last 10\% of the run.

For the pore-sealing simulations, the pore area was computed on a grid of
$0.5 \times 0.5$~nm$^2$ columns perpendicular to the bilayer plane. A column is empty when it
contains no bead, and the pore is the set of connected empty regions that reach within
1.5~nm of the pore center; isolated empty cells are gaps between beads and were ignored. The
pore center is the center of the largest empty region in the first frame, moved with the mean
lateral drift of all beads. The pore radius is $\sqrt{A/\pi}$, with $A$ the pore area
(Tables~\ref{tab:complex-systems} and~\ref{tab:dopcdppc}), and the sealing time is the first
time after which the pore area is zero in at least 98\% of the next 50 frames, computed from
trajectories saved every 0.16~cg-ps. The PO4 roughness in Table~\ref{tab:complex-metrics} is the
standard deviation of the PO4 heights of the upper leaflet. Same as for the curved systems
the local normal of a lipid, used for the values marked with $\dagger$, is the mean director, 
oriented towards $+z$, of all lipids within 2~nm of it in the bilayer plane.

\section{Sensitivity and robustness studies}\label{sec:si-sensitivity}
Beyond the performance of the \texttt{Lipid-MLCG} model across several
conformations, we also assessed its robustness and reproducibility. To this end,
we performed an ablation of the prior terms, together with sensitivity and
variability studies, in which we trained
models under different conditions and evaluated whether they stabilize the
150-lipid bilayer presented in the main text (section ``Bilayer structure and dynamics'').
\subsection{Prior ablation}
\begin{figure}
    \centering
    \includegraphics[width=\linewidth]{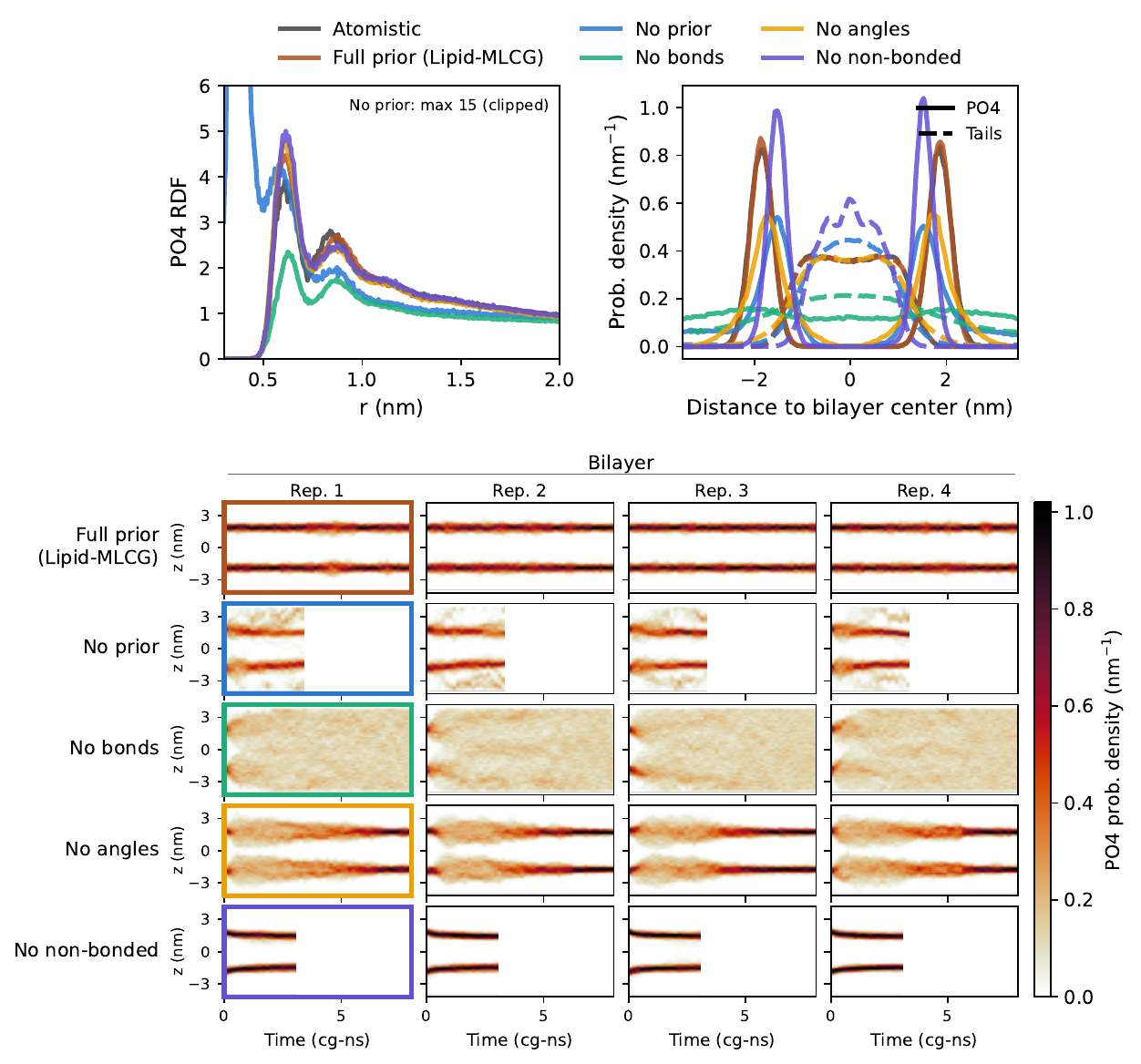}
    \caption{Prior ablation study: models trained without a prior and without each individual
    prior term, tested by simulating a small bilayer.
    (Middle) headgroup radial distribution function and headgroup (solid) and tail-bead 
    (dashed) density profiles along the bilayer normal, combining all replicas, compared 
    with the atomistic reference (gray).
    (Bottom) headgroup density along $z$ as a function of time for each replica.}
    \label{fig:prior-ablation}
\end{figure}

\begin{figure}
    \centering
    \includegraphics[width=0.8\linewidth]{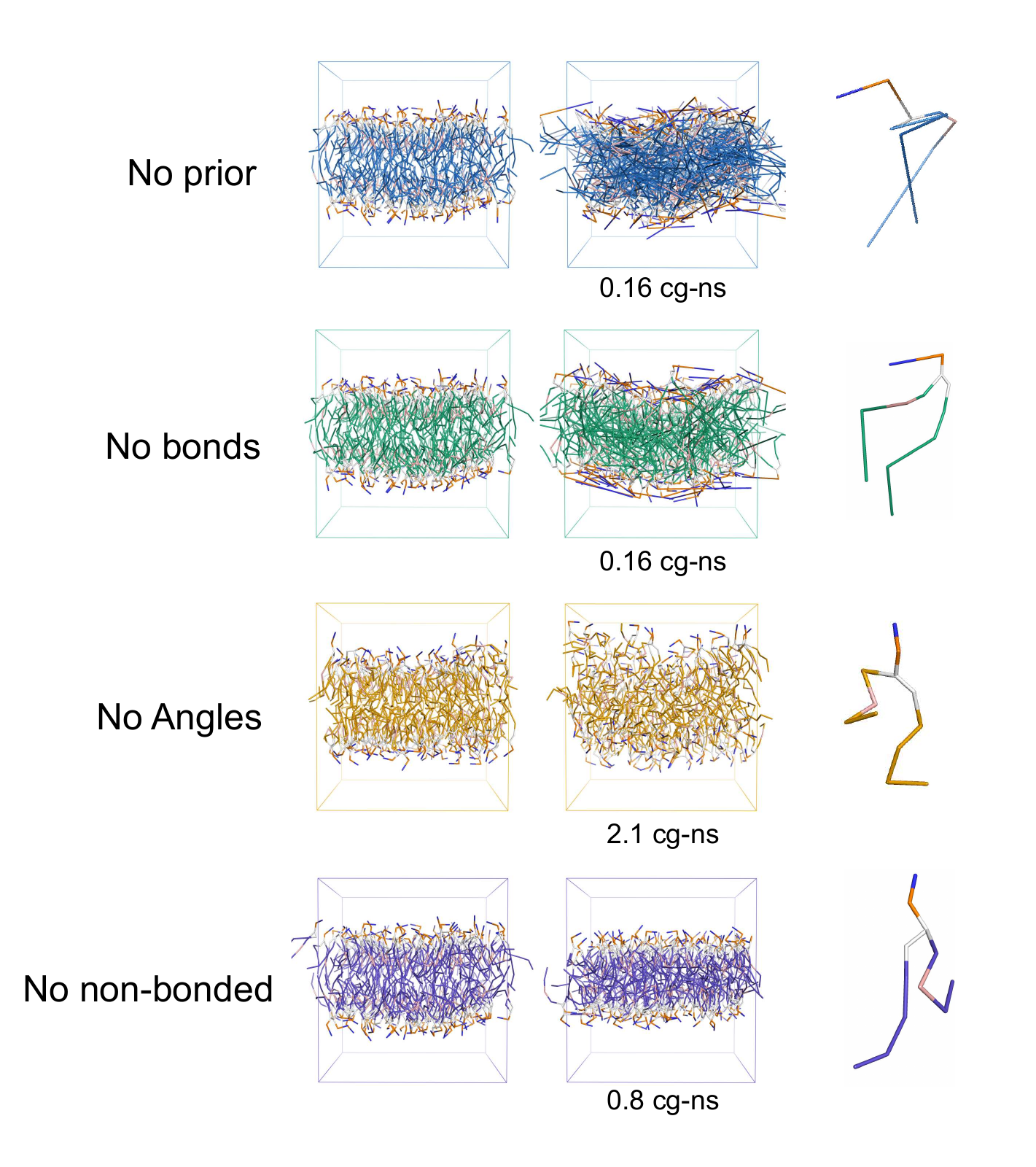}
    \caption{(Left, Middle) Structure snapshots and (Right) single lipid selections for the models of figure
    ~\ref{fig:prior-ablation}}
    \label{fig:prior-ablation-details}
\end{figure}

First, we probed the relevance of each prior term by ablating the entire prior
and each of its terms individually (Figure~\ref{fig:prior-ablation}).
The models without a prior and without bond terms cannot keep individual lipid
molecules stable, let alone the entire bilayer. The model without non-bonded terms
preserves the bilayer structure but compresses the lipids together. The model
without angle terms appears to resemble the full-prior model, but closer
inspection reveals that the tail-end beads bend back toward the headgroup,
leading to unphysical contacts between tail beads (Figure~\ref{fig:prior-ablation-details}).
\subsection{Network cutoff}
\begin{figure}
    \centering
    \includegraphics[width=\linewidth]{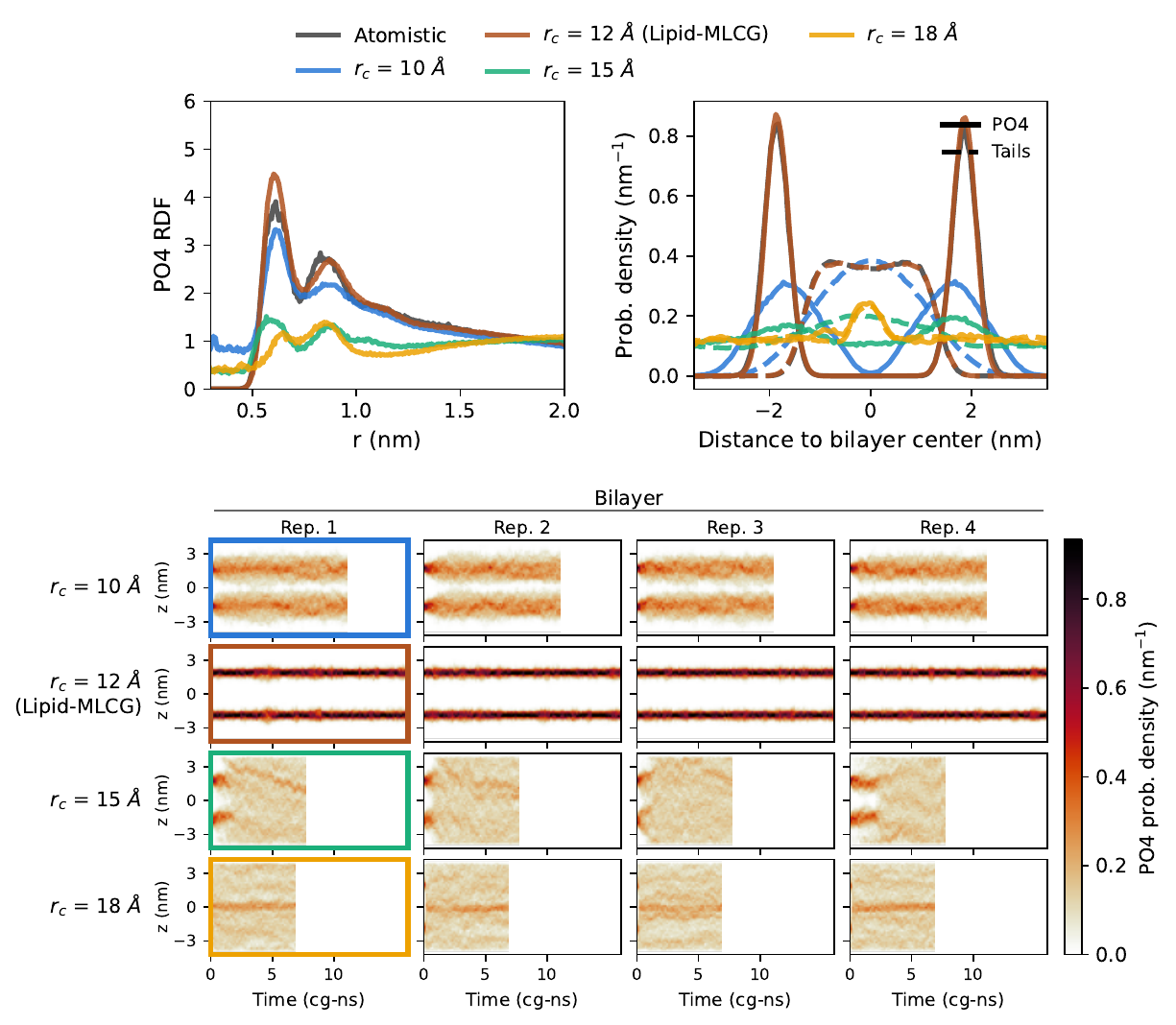}
    \caption{Sensitivity to the network cutoff: models trained with cutoffs of 10, 12, 15 and 18~\AA{}.
    Panels follow the same distribution as figure~\ref{fig:prior-ablation}}
    \label{fig:cutoff-ablation}
\end{figure}

Next, we assessed the effect of the network cutoff by training models with
cutoffs of 10, 12, 15 and 18~\AA{} (Figure~\ref{fig:cutoff-ablation}).
While the models with 10 and 12~\AA{} cutoffs stabilize the bilayer, those with
larger cutoffs do not. We attribute this to the fixed number of radial basis
functions: a larger cutoff decreases the number of basis functions per unit
distance and therefore the radial resolution of the network.
\subsection{Minibatch composition}
\begin{figure}
    \centering
    \includegraphics[width=\linewidth]{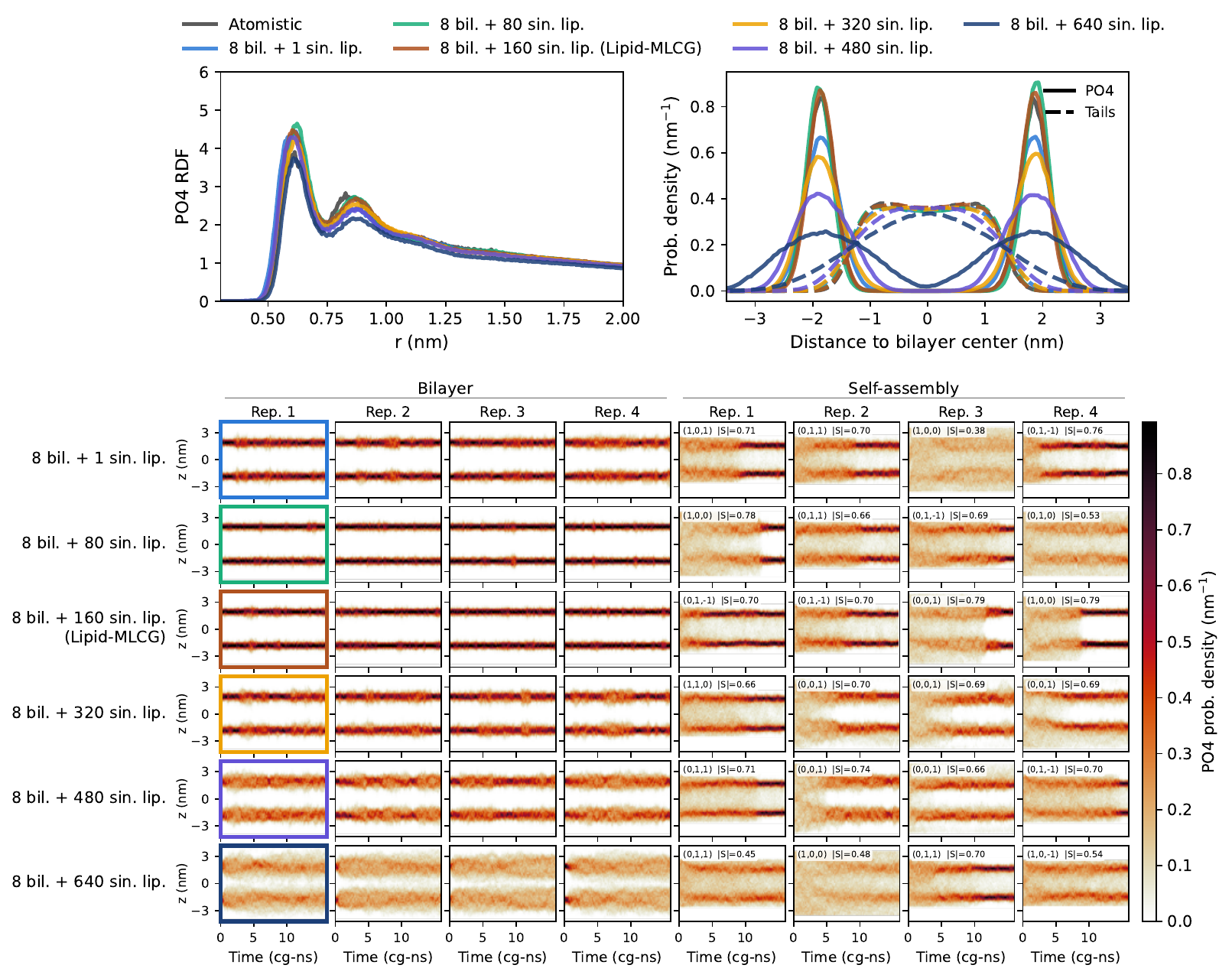}
    \caption{Sensitivity to the minibatch composition: bilayer stability and self-assembly for
    models trained with 1, 80, 160 (\texttt{Lipid-MLCG}) and 320 single-lipid frames per
    minibatch, each with 8 bilayer frames.
    Panels follow the same distribution as figure~\ref{fig:prior-ablation}}
    \label{fig:dataset-ablation}
\end{figure}

Because the two training systems differ greatly in size, the ratio of single-lipid 
to bilayer frames in each minibatch determines how strongly each system shapes the learned forces. 
To test its influence, we kept the 8 bilayer frames per minibatch of \texttt{Lipid-MLCG} and 
varied the number of single-lipid frames between 1, 80, 160 and 320 (Figure~\ref{fig:dataset-ablation}). 
The bilayer remained stable for every composition, and most replicas also self-assembled from dispersed lipids. 
The headgroup distribution, however, proved more sensitive: with 1 or 320 single-lipid frames the $z$-density 
peaks broaden and lose height relative to the atomistic reference, whereas intermediate compositions 
(80 and 160 frames) reproduce it closely. Bilayer stability is thus robust to the minibatch composition, 
while the structural accuracy of the bilayer benefits from a balanced mixture of the two systems.
\subsection{Training noise level}
\begin{figure}
    \centering
    \includegraphics[width=\linewidth]{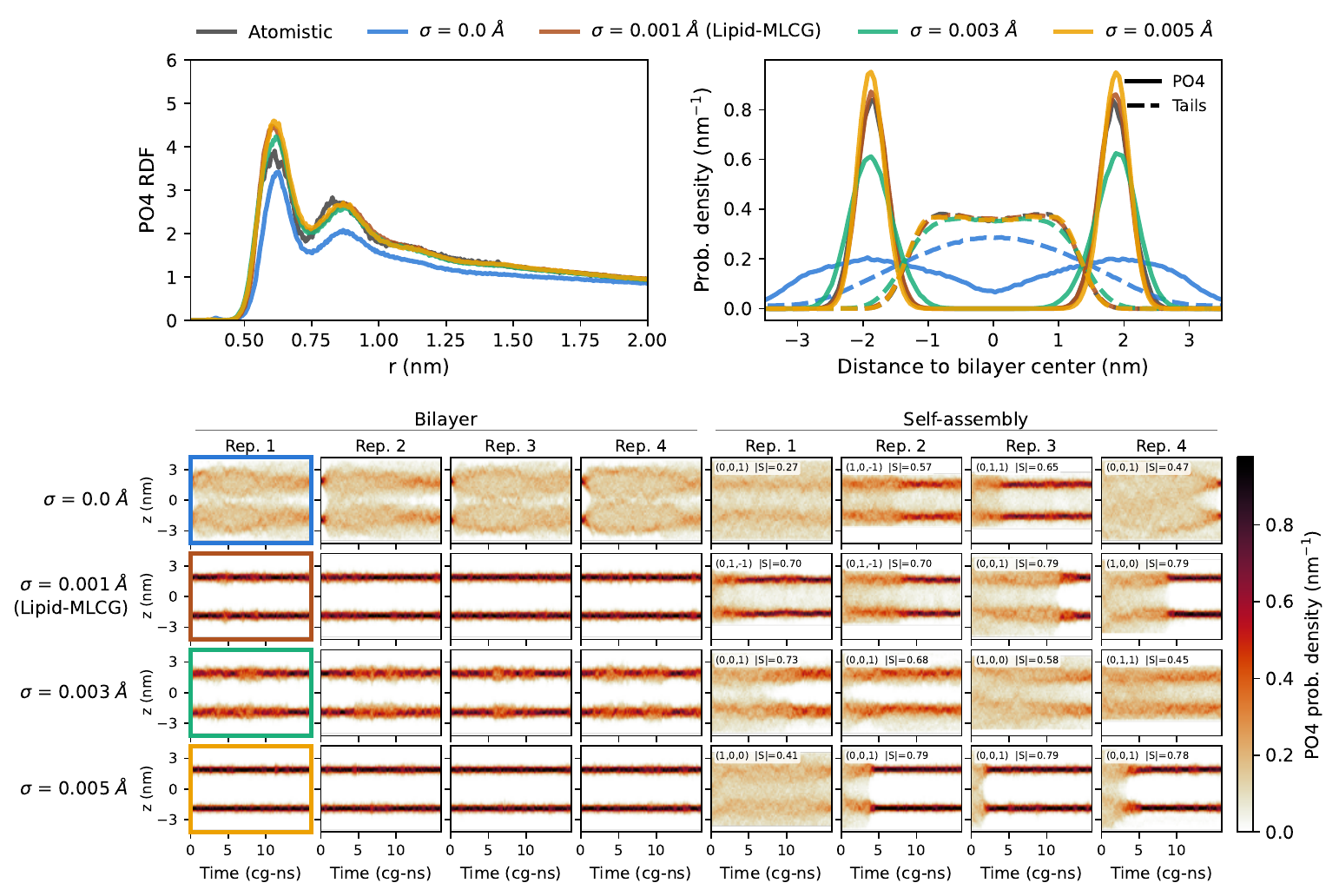}
    \caption{Sensitivity to the training noise: bilayer stability and self-assembly for models trained with
    $\sigma = 0$, 0.001, 0.003 and 0.005~\AA{}. Panels follow the same distribution as figure~\ref{fig:prior-ablation}}
    \label{fig:noise-ablation}
\end{figure}

We also varied the training noise level. In addition to
the main model ($\sigma = 0.001$~\AA{}), we trained models without noise and with
$\sigma = 0.003$ and $0.005$~\AA{}. The model without noise did not stabilize the
bilayer, whereas all models trained with noise both stabilized the bilayer and
self-assembled it from dispersed lipids (Figure~\ref{fig:noise-ablation}). However, while the model with
$\sigma = 0.001$~\AA{} closely matched the reference headgroup $z$-density
profile, the models with higher noise over-stabilized the bilayer. We therefore
chose $\sigma = 0.001$~\AA{}.
\subsection{Random-seed variability}
\begin{figure}[h]
    \centering
    \includegraphics[width=\linewidth]{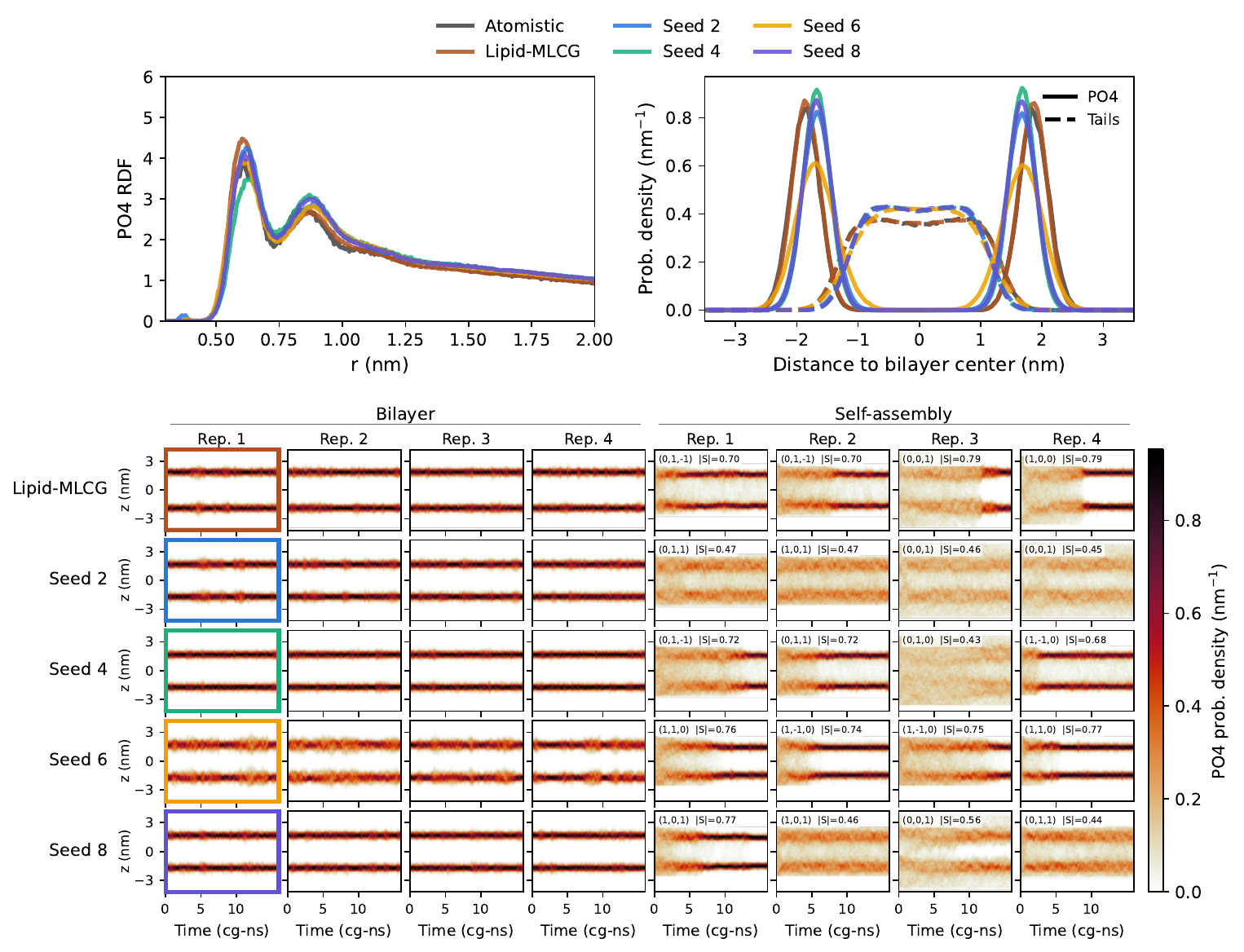}
    \caption{Random-seed variability: bilayer stability and self-assembly for four additional
    models initialized with different random seeds.Panels follow the same 
    distribution as figure~\ref{fig:prior-ablation}}
    \label{fig:seed-ablation}
\end{figure}

Finally, we evaluated the variability across random seeds. We trained and
simulated four additional models, initialized with random seeds different from
that of the model presented in the main text. All four models stabilized the
bilayer, and three of them also self-assembled a flat bilayer (Figure~\ref{fig:seed-ablation}).

\newpage
\section{Supporting figures}
\begin{figure}[h]
    \centering
    \includegraphics[width=0.9\linewidth]{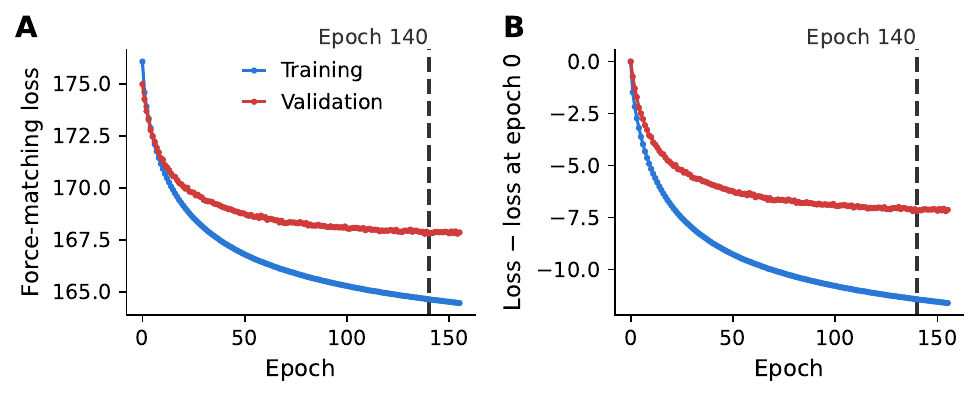}
    \caption{(A) Absolute and (B) relative force-matching training (blue) and validation (red) losses versus epoch for the
    \texttt{SinLip-MLCG} model. The black dotted line shows the selected simulation epoch of figure.}
    \label{fig:single_lipid_train_curve}
\end{figure}
\begin{figure}[h]
    \centering
    \includegraphics[width=0.9\linewidth]{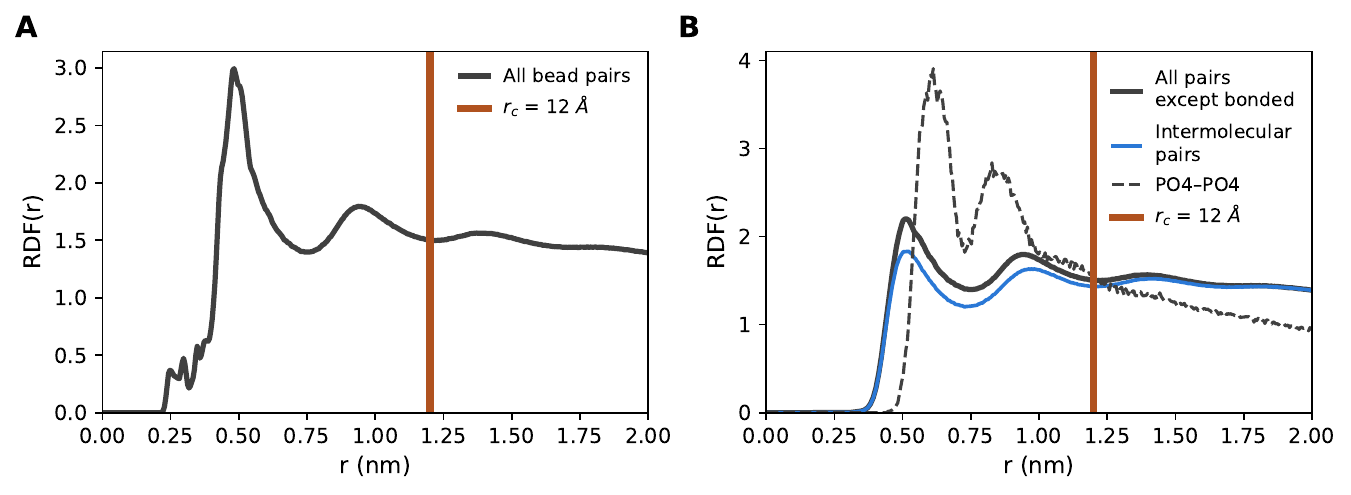}
    \caption{Radial distribution function of the CG beads in the mapped atomistic bilayer. (A) shows the 
    RDF of all beads in the system, while (B) shows RDFs excluding bonded pairs, intramolecular interactions 
    and using only headgroups pairs. In all cases, the first non-trivial two peaks are covered by the 12 \AA{}~cutoff}
    \label{fig:RDF}
\end{figure}
\begin{figure}
    \centering
    \includegraphics[width=0.98\linewidth]{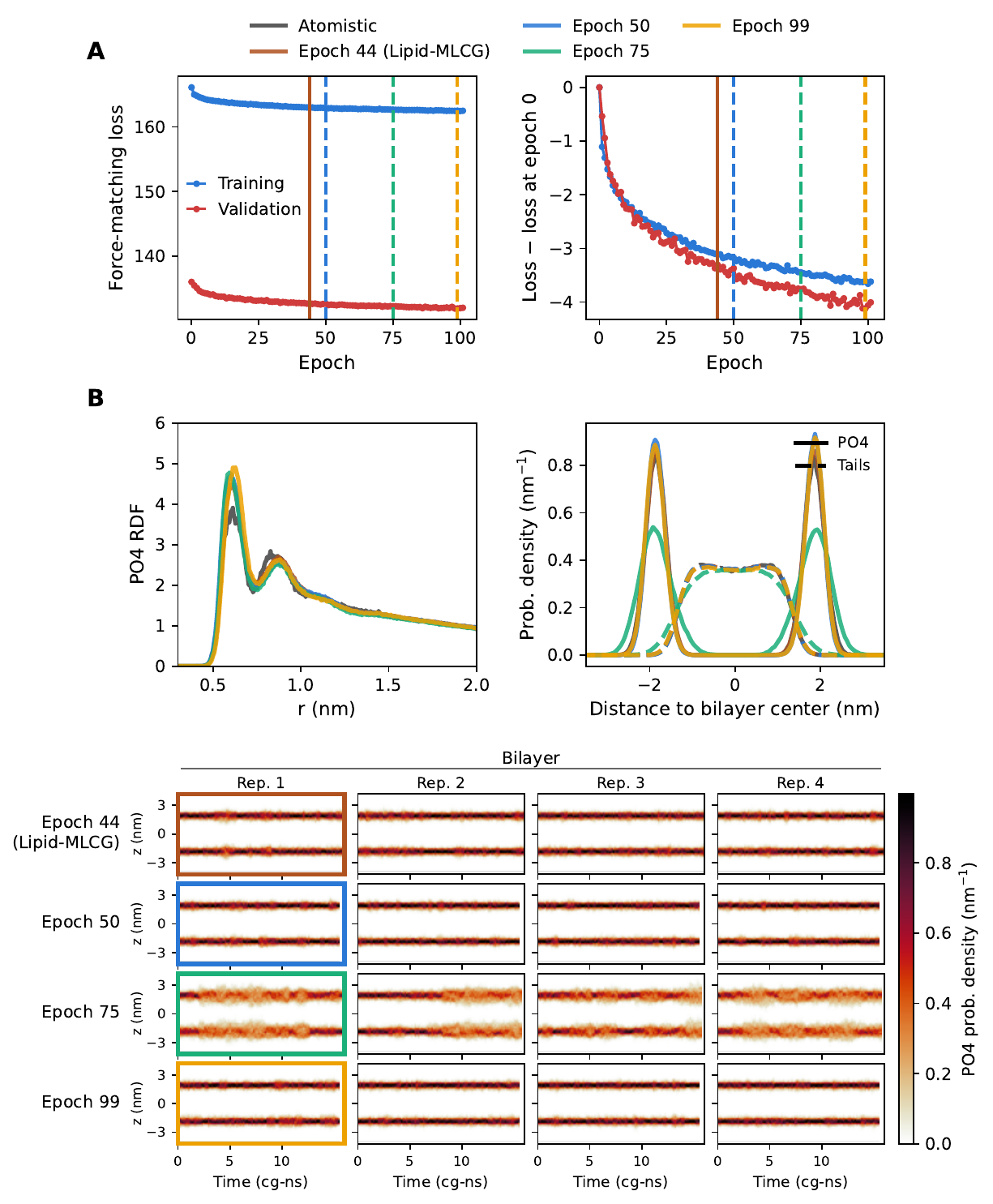}
    \caption{Epoch selection of the \texttt{Lipid-MLCG} model.(A) (left) Absolute and (right) relative training and validation losses. The vertical lines
    show the set of epochs chosen for evaluation. (B) Simulation of a 150-lipid bilayer with the selected epochs. While epoch 44, 50 and 99 are capable 
    of keeping the bilayer in place, epoch 44 has presents a slightly better  match in the RDF.}
    \label{fig:epoch_selection}
\end{figure}
\begin{figure}
    \centering
    \includegraphics[width=\linewidth]{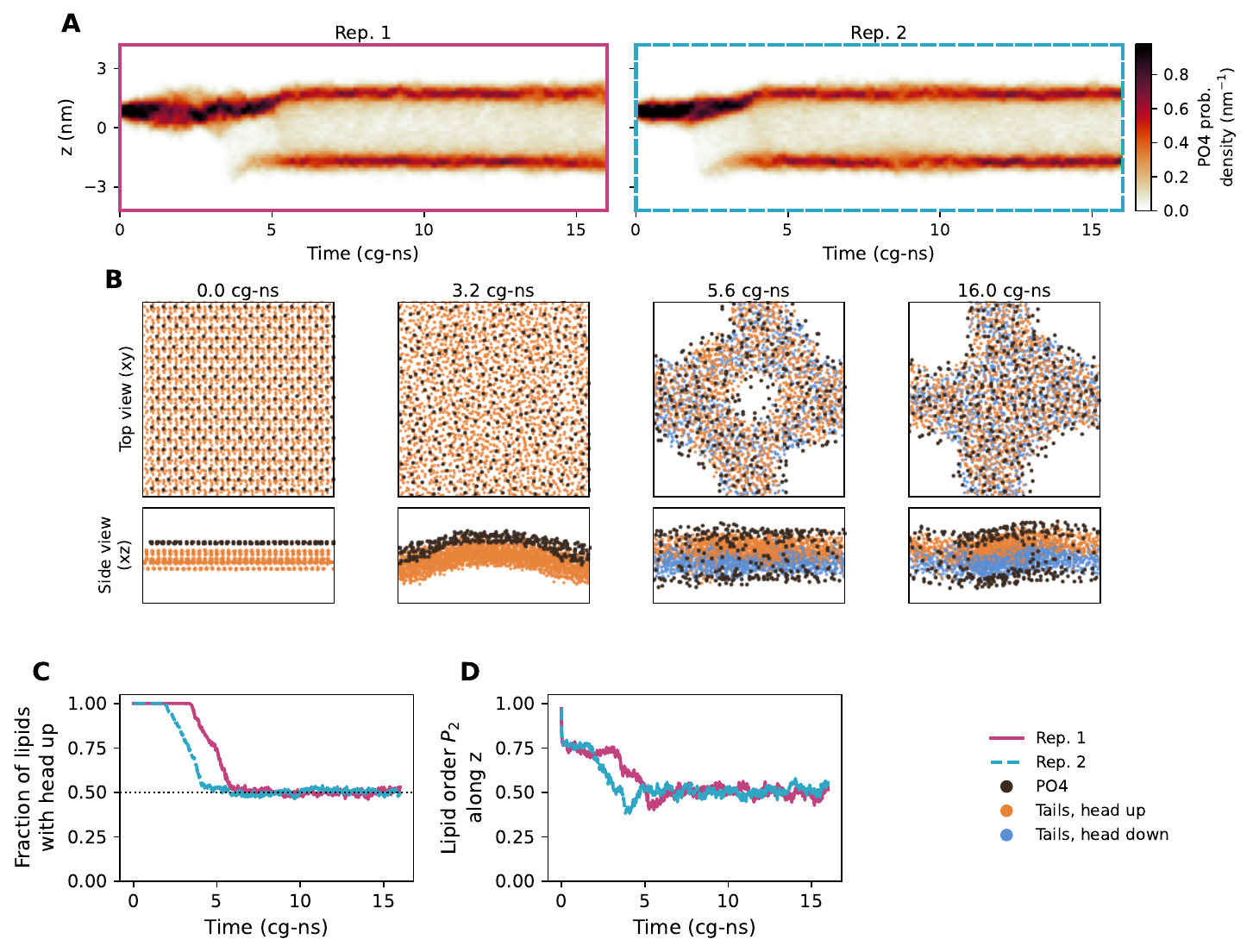}
    \caption{Simulation of a monolayer, showing the emergence of a bilayer over time. (A) z density of PO4 heads over time.
    (B) top and side view snapshots from the simulation, showing the emergence of the bilayer after 5 cg-ns. (C) 
    fraction of lipids with head up and (B) lipid alignment order parameter $P_2$ as function of simulation time. }
    \label{fig:monolayer}
\end{figure}
\begin{figure}
    \centering
    \includegraphics[width=\linewidth]{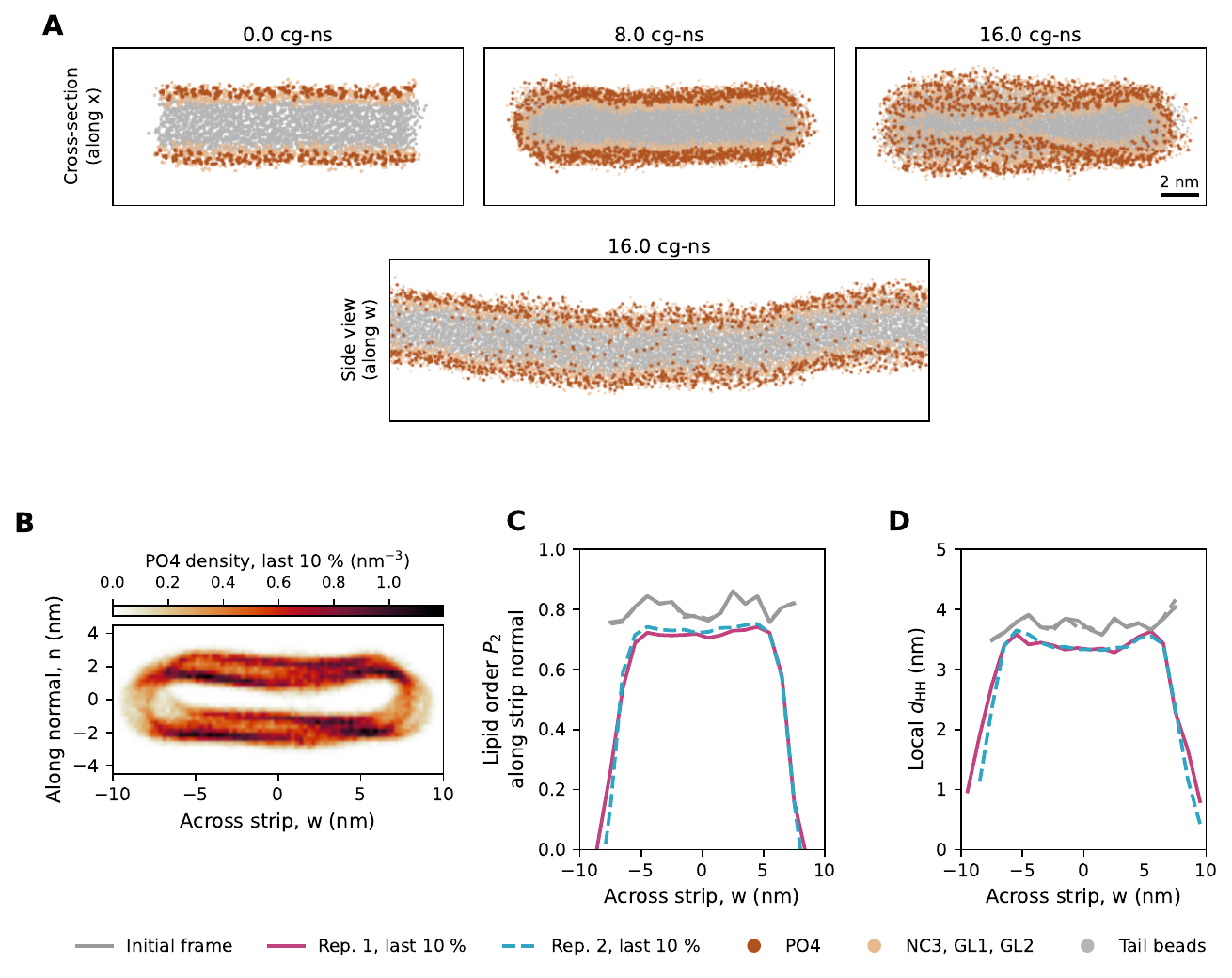}
    \caption{Simulations of a bilayer strip that spans the periodic box along $x$ but leaves free
    space along $y$, showing the formation of rounded, micelle-like edges. (A) Cross-section along 
    $x$ at the initial, middle and end of the Replica 1, with a side view of the end.
    The snapshot shows the lipids moving across the cross section to wrap the tails 
    shielding them from the solvent space. (B) PO4 density of Replica 1 in the cross-section of the strip, 
    as a function of the position across the strip ($w$) and along its normal ($n$), averaged 
    over the last 10\% of the run and along $x$. (C) Lipid alignment order with  $P_2$ with respect 
    to the strip normal and (D) local $d_{HH}$ (as the difference between the median PO4 
    heights of the two leaflets, which limits the effect of the lipids at the edges 
    across the strip direction). $P_2$ and $d_{HH}$ are plotted in the initial frame
    and over the last 10\% of each replica. At the edges, the lipids point outwards along the rim,
so $P_2$ with respect to the strip normal drops towards zero.}
    \label{fig:wrongboundary}
\end{figure}
\begin{figure}
    \centering
    \includegraphics[width=0.9\linewidth]{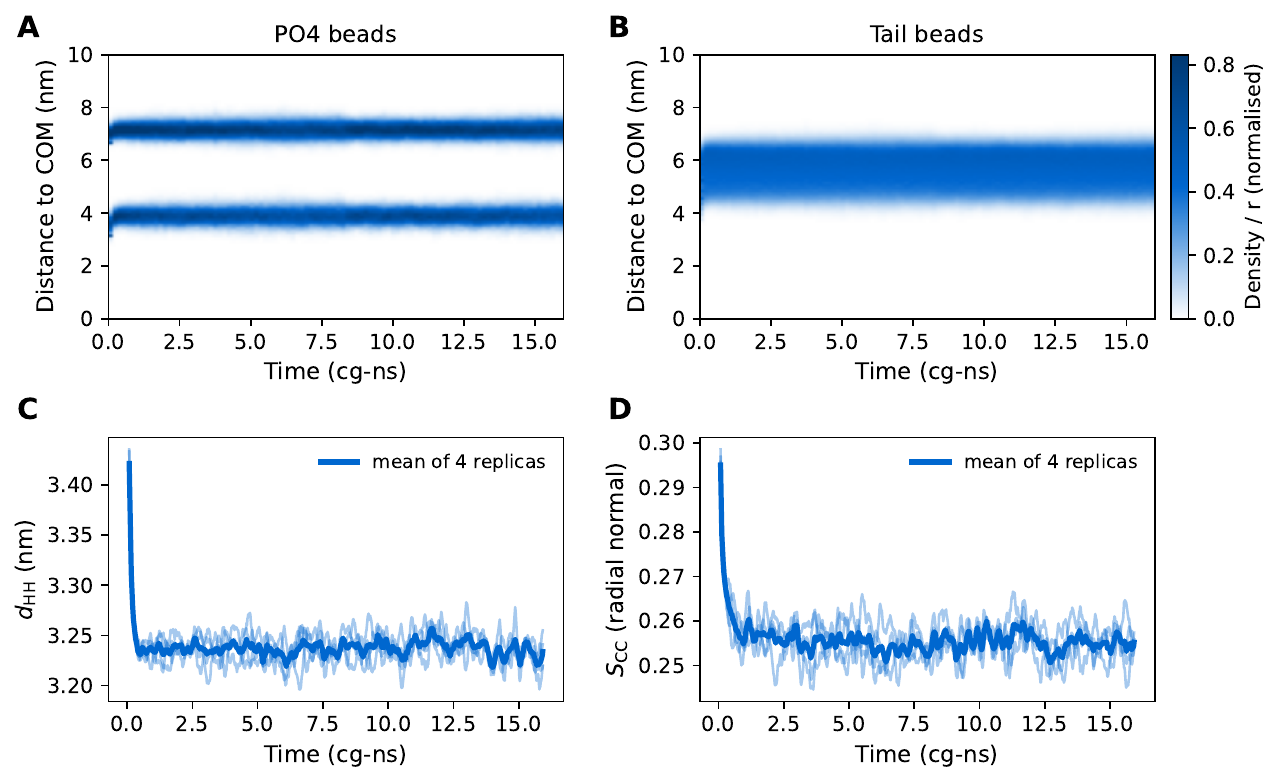}
    \caption{Time evolution of the small vesicle (4 independent runs). (A) PO4 and (B)
    tail-bead density as a function of the distance to the center of mass and of time, combining
    all runs, in windows of 0.16~cg-ns. The histograms are divided by $r$ and normalized. 
    (C) $d_{HH}$, the difference between the median PO4 radii of the two leaflets, and 
    (D) $S_{CC}$ with respect to the radial direction, as a function of time. 
    Light lines show the individual runs and thick lines their mean (running mean over
    0.16~cg-ns).}
    \label{fig:small_vesicle_time}
\end{figure}
\begin{figure}
    \centering
    \includegraphics[width=0.9\linewidth]{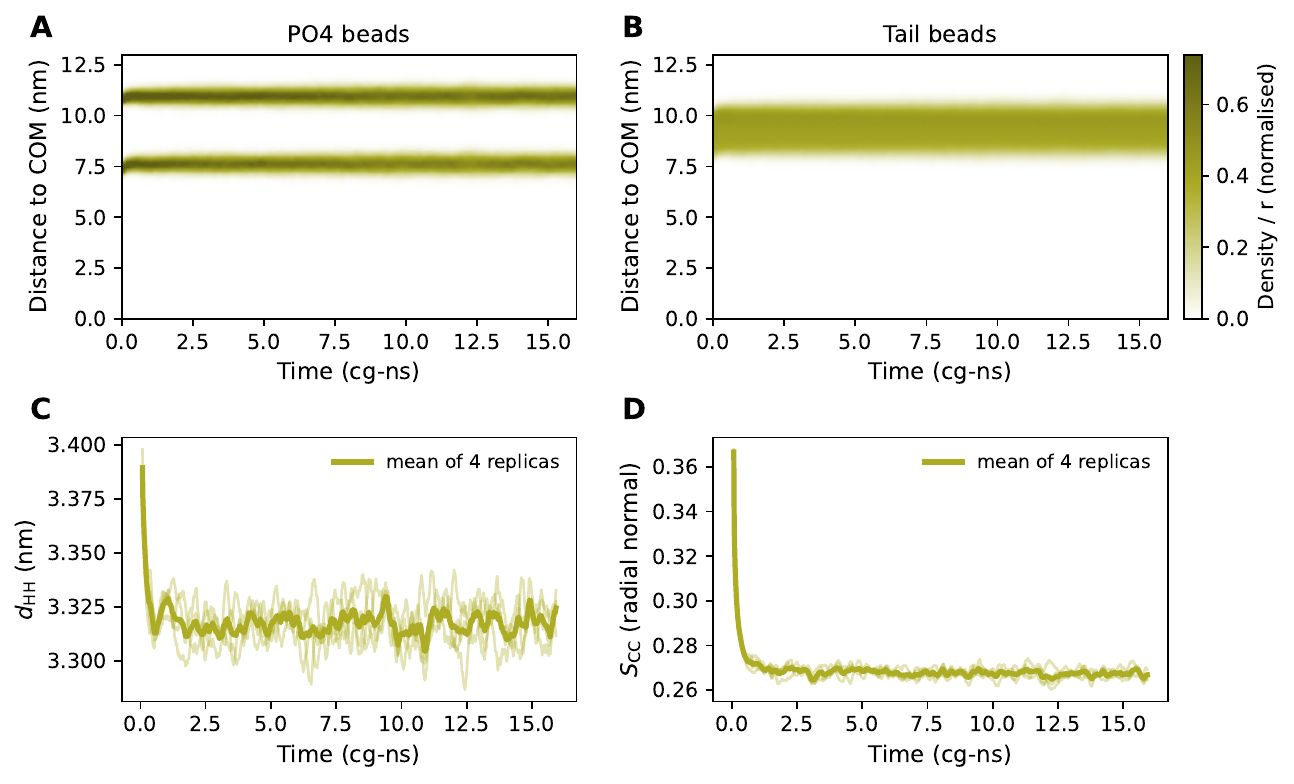}
    \caption{Time evolution of the large vesicle (4 independent runs). Panels and calculations 
    are the same as in Figure~\ref{fig:small_vesicle_time}}
    \label{fig:large_vesicle_time}
\end{figure}
\begin{figure}
    \centering
    \includegraphics[width=0.9\linewidth]{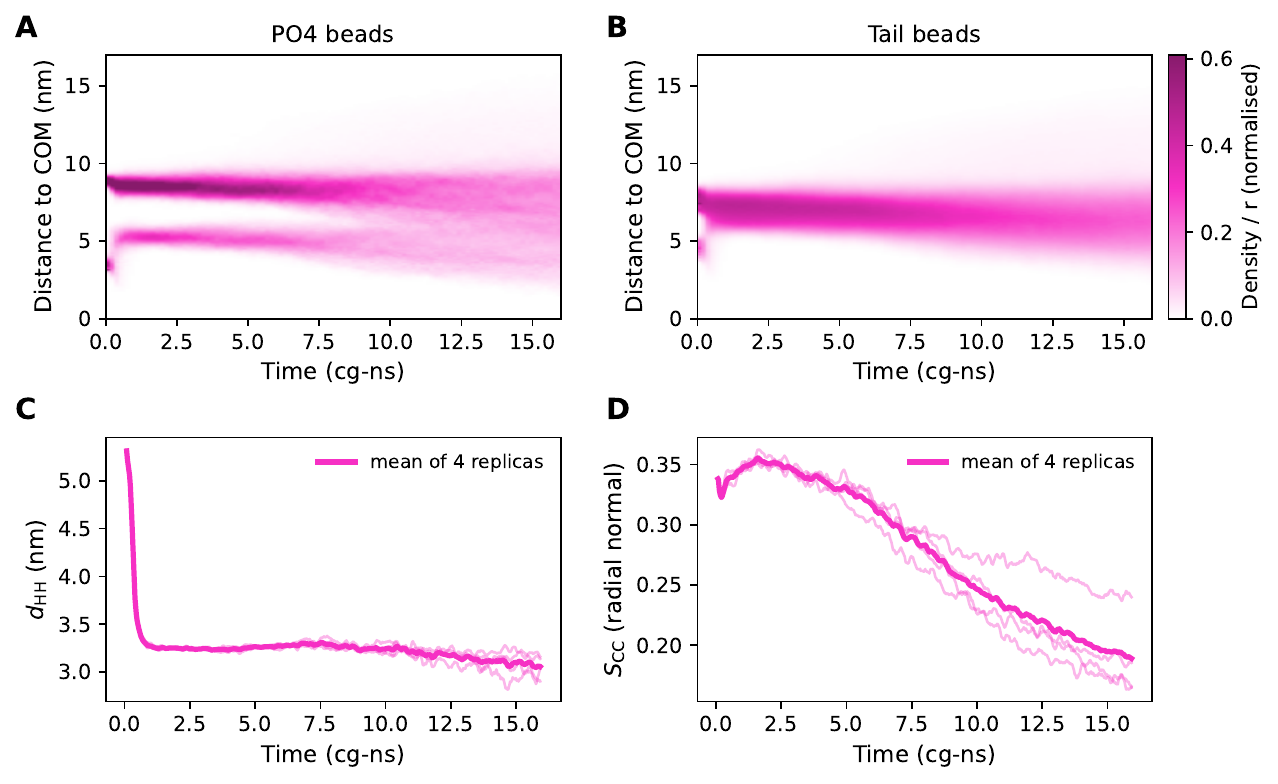}
    \caption{Time evolution of the wrongly shaped vesicle (4 independent runs). Panels and calculations 
    are the same as in Figure~\ref{fig:small_vesicle_time}}
    \label{fig:wrong_vesicle_time}
\end{figure}
\begin{figure}
    \centering
    \includegraphics[width=0.9\linewidth]{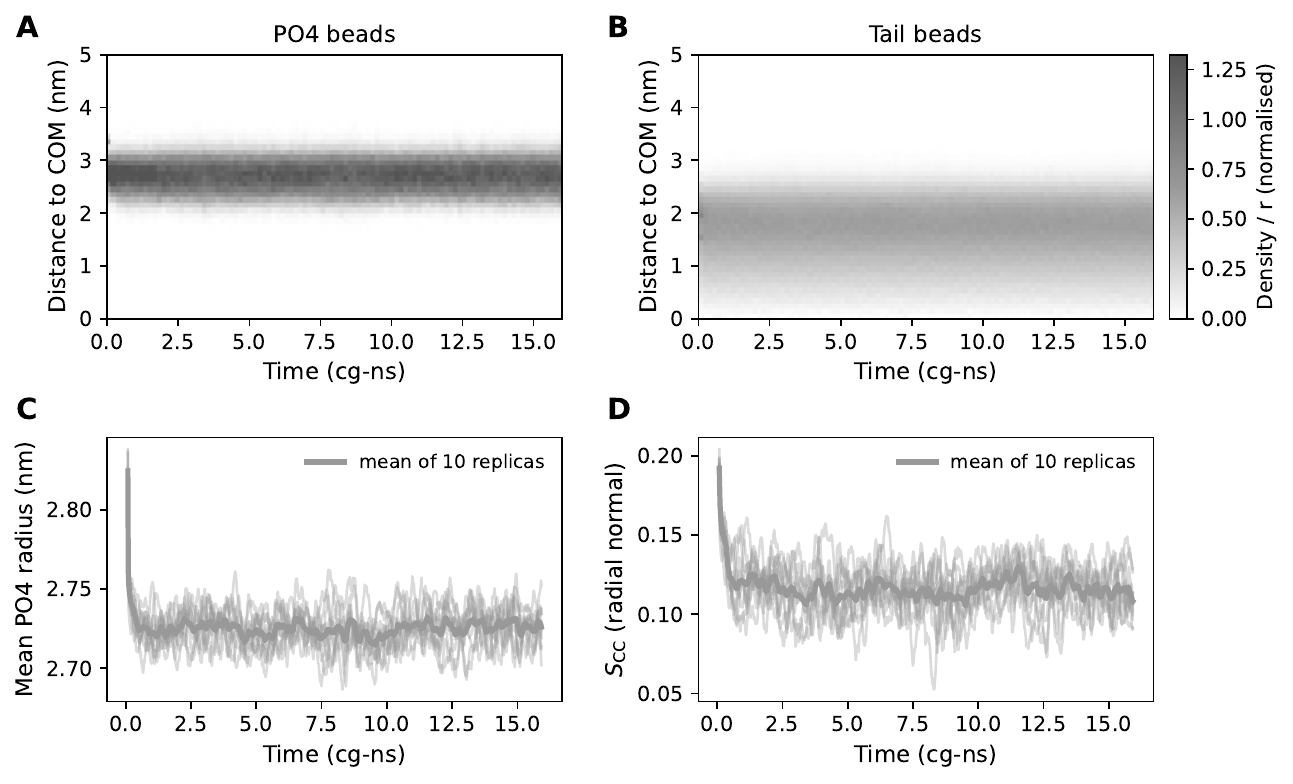}
    \caption{Time evolution of the wrongly shaped vesicle (10 independent runs). Panels and calculations 
    are the same as in Figure~\ref{fig:small_vesicle_time}, with exception of panel (C) 
    which shows the mean PO4 distance to the center of mass instead of $d_{HH}$, since the
    micelle has a single layer.}
    \label{fig:micell_time}
\end{figure}
\begin{figure}
    \centering
    \includegraphics[width=\linewidth]{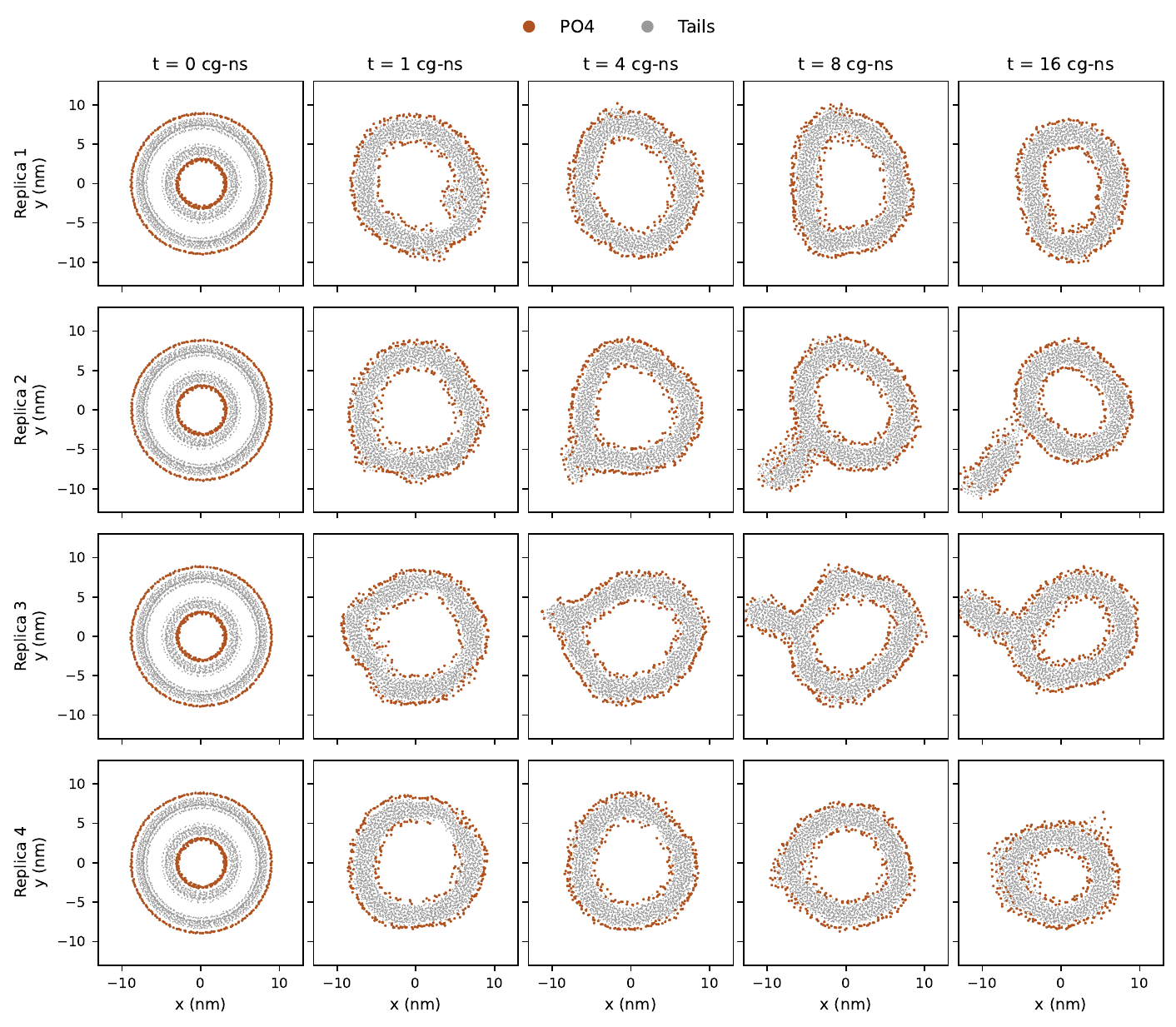}
    \caption{Snapshots along a the four independent replicas  of the wrongly shaped vesicle, showing the formation of
    micelle-like bulges attached to the outer leaflet.}
    \label{fig:wrong_vesicle_detail}
\end{figure}
\begin{figure}
    \centering
    \includegraphics[width=0.95\linewidth]{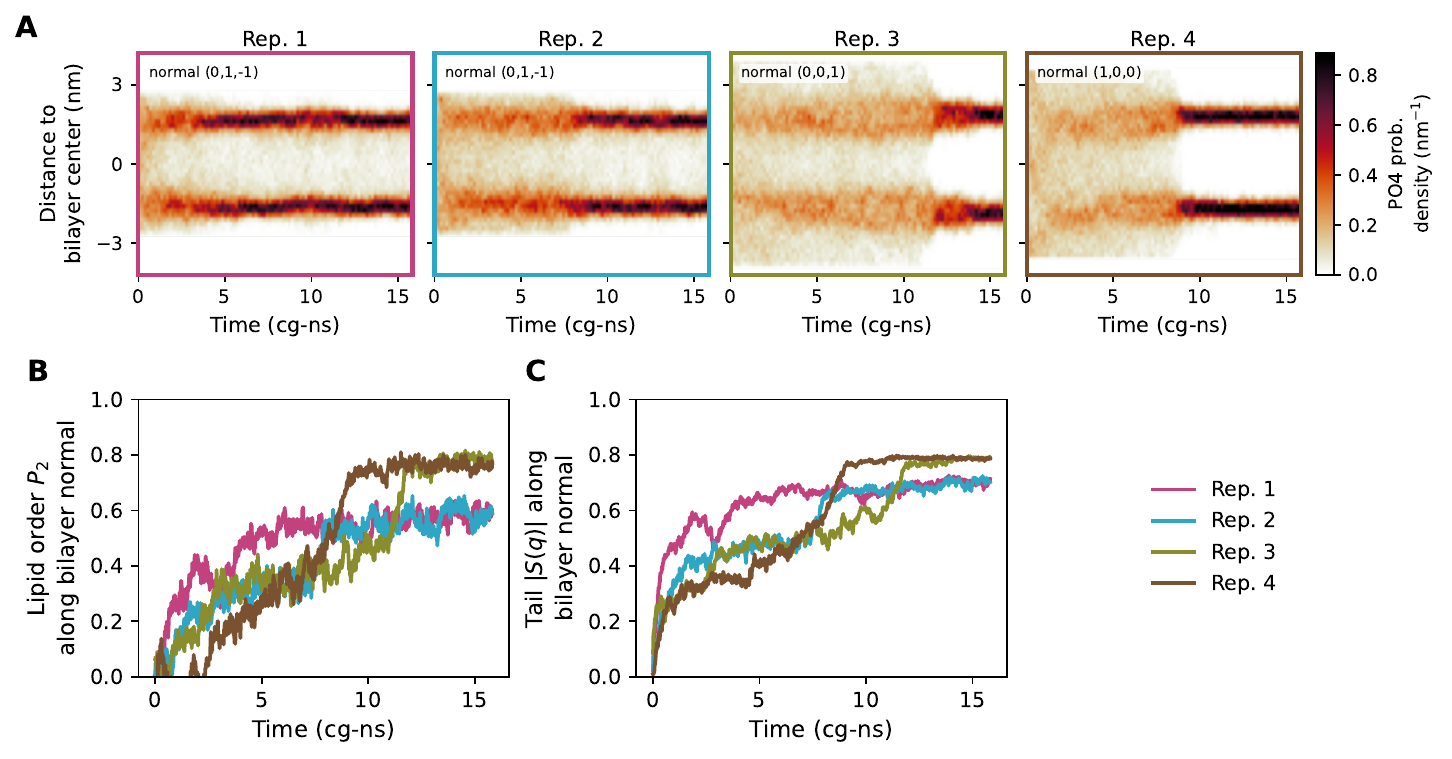}
    \caption{Details on the bilayer self-assembly from the high-density, uniformly distributed system over four
    replicas. (A) Heatmap of the headgroup density as function of the simulation time. (B) Lipid ordering 
    parameter $P_2$ along the bilayer normal (C) Structure factor $\vert S(q) \vert$ of the carbon tails.}
    \label{fig:bilayer_self_assembly}
\end{figure}
\begin{figure}
    \centering
    \includegraphics[width=0.95\linewidth]{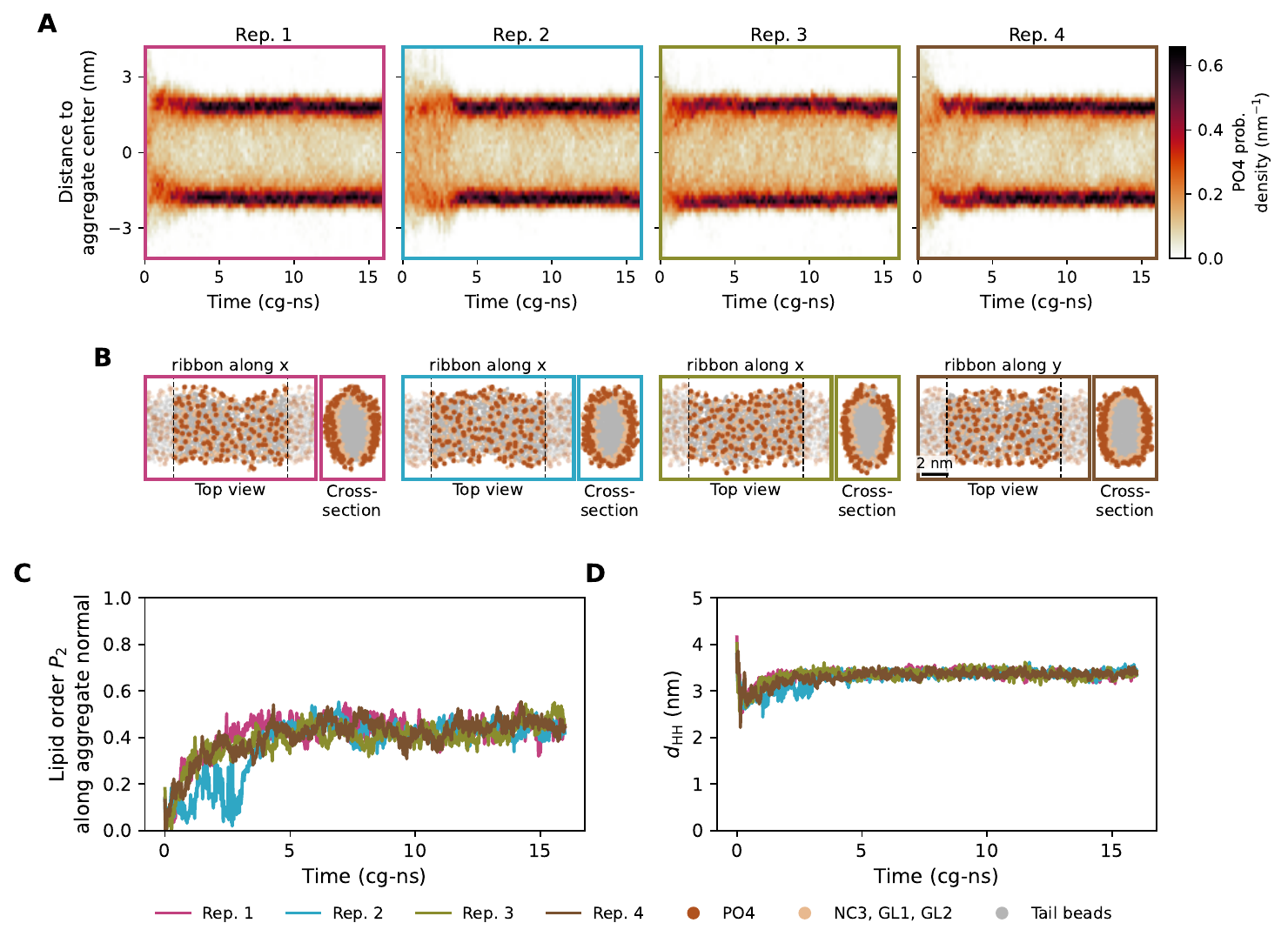}
    \caption{Self-assembly of a bicelle from a low-density, uniformly distributed system.
    (A) Heatmap of the headgroup density as function of the simulation time for 4 replicas. 
    (B) snapshots of the simulation along the top and cross view of a selected formed ribbon. 
    (C) Lipid ordering parameters $P_2$ along the aggregate normal and (D) $d_{HH}$ the bilayer thickness, 
    defined as the difference between the median PO4 heights of the two leaflets,
    which limits the effect of the lipids at the rims.}
    \label{fig:bilayer_wrong_self_assembly}
\end{figure}
\begin{figure}
    \centering
    \includegraphics[width=0.95\linewidth]{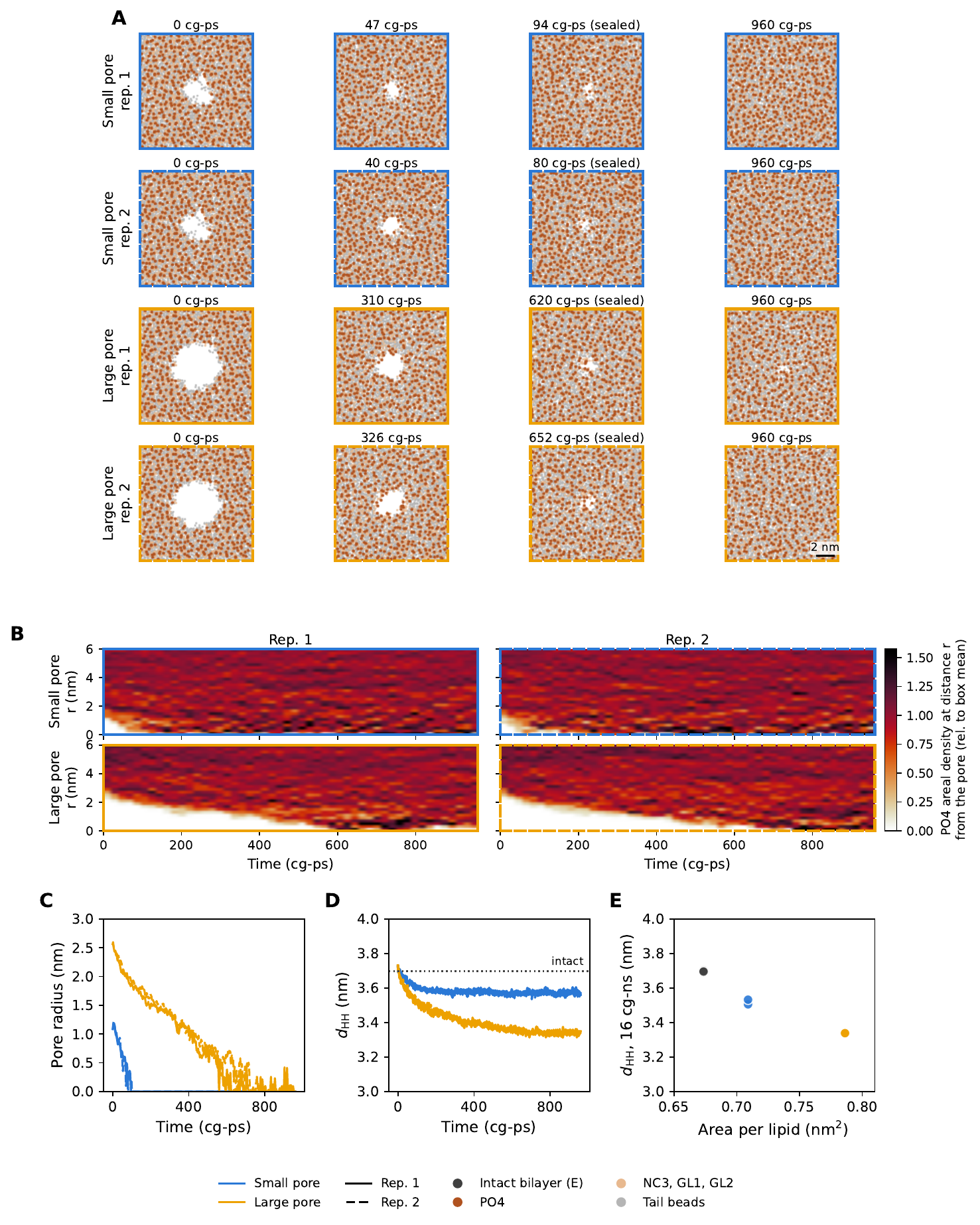}
    \caption{Sealing of the small and large pores, in two independent trajectories for each
    system.
    (A) Trajectory snapshots, from the top view, for the 2 replicas in each system, showing the closing of the pore
    (B) Heatmap of headgroup areal density from the box mean, showing the closure of the pore. 
    (C) Pore radius as function of the simulation time 
    (D) $d_{HH}$ as function of the simulation time. Dashed line shows the $d_{HH}$ of the intact bilayer.
    (E) Comparison of bilayer thickness and area per lipid at the end of the simulation for the intact bilayer, 
    the small pore and the large pore. See section \ref{sec:density-along-bilayer-normal} for details on 
    the calculated quantities}
    \label{fig:pore_closing}
\end{figure}
\begin{figure}
    \centering
    \includegraphics[width=\linewidth]{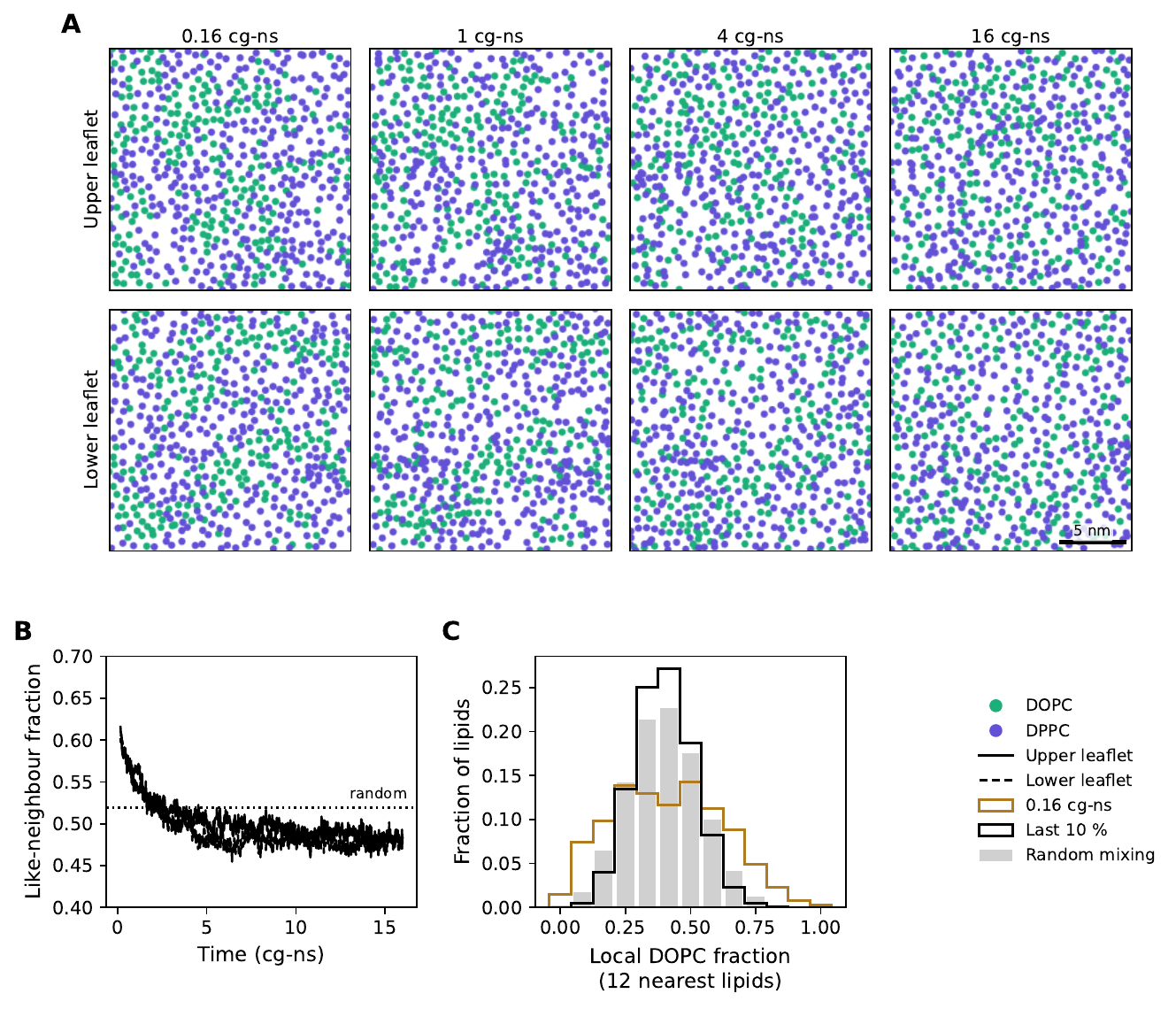}
    \caption{Lateral distribution of DOPC (green) and DPPC (purple) in the mixed bilayer at 300 K
    (460 DOPC + 692 DPPC), started from a partially separated configuration. (A) Top views of each leaflet. 
    (B) Fraction of like neighbours (6 nearest lipids in the same leaflet); dotted: random mixing. 
    (C) Local DOPC fraction around each lipid compared with random mixing. The lipids mix within 
    $\sim$4~cg-ns and no domains form.}
    \label{fig:dppcdopc_no_aggr}
\end{figure}
\begin{figure}
    \centering
    \includegraphics[width=\linewidth]{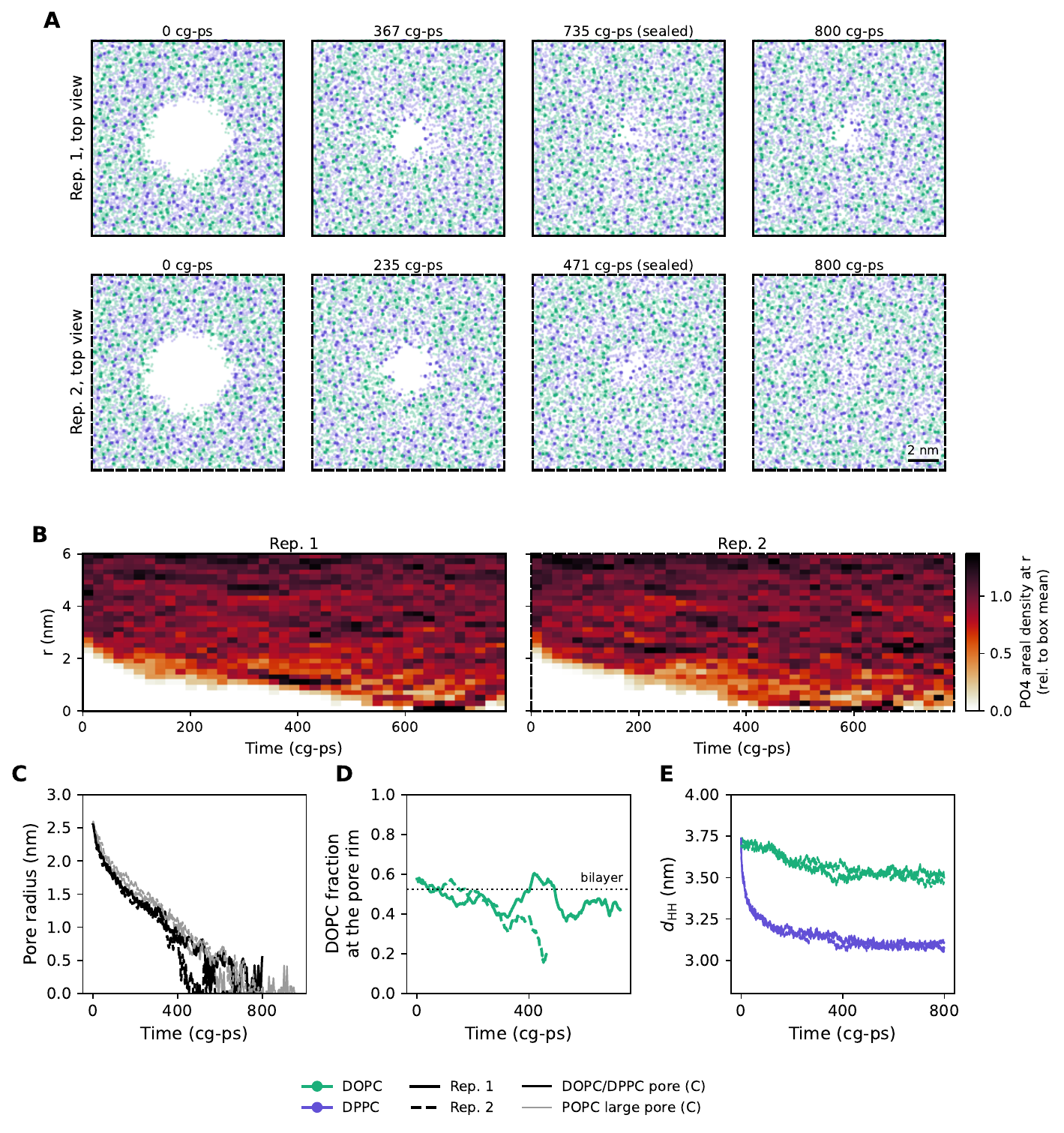}
    \caption{Pore sealing in the DOPC/DPPC bilayer.
    (A) Snapshots of the simulations at different time, showing th sealing across two replicas.
    (B) Headgroup areal density as a function of the distance $r$ to the pore center and of time, relative
    to the mean density of the box.
    (C) Pore radius as a function of time (running mean over 4~cg-ps). Gray lines show the large 
    POPC pore, which has the same number of lipids, box and initial radius.
    (D) DOPC fraction among the lipids at the pore rim (headgroup within 1.2~nm of the pore edge) while the
    pore is open (running mean over 16~cg-ps). The dotted line shows the DOPC fraction of the whole
    bilayer.
    (E) Bilayer thickness $d_{HH}$ computed separately for DOPC and DPPC lipids, as a function of
    time.
    }
    \label{fig:dppcdopc_hole_details}
\end{figure}
\clearpage
\printbibliography
\end{document}